\documentclass[12pt]{article}

\usepackage{times}
\usepackage{mdframed}

\usepackage{amsthm,amsmath}
\usepackage{xcolor,colortbl}
\usepackage{array}
\usepackage[pdftex]{graphicx}   
\usepackage{framed}
\usepackage[font={small}]{caption}
\usepackage{lineno}

\newenvironment{sciabstract}{%
\begin{quote} }
{\end{quote}}

\newcounter{lastnote}

\title{What can ecosystems learn? \\Expanding evolutionary ecology with learning theory.}

\author
{Daniel A. Power,$^{1\ast}$ Richard A. Watson,$^{1,2}$, E\"ors Szathm\'ary$^{3}$, Rob Mills$^{4}$,\\ Simon T Powers$^{5}$, C Patrick Doncaster$^{6}$ and B\l a\.zej Czapp$^{6}$ \\
\\
\normalsize{$^{1}$Electronics and Computer Science, University of Southampton, UK} \\
\normalsize{$^{2}$Institute for Life Science, University of Southampton, UK}\\
\normalsize{$^{3}$The Parmenides Center for the Conceptual Foundations of Science, Pullach, Germany}\\
\normalsize{$^{4}$Department of Informatics, Faculty of Sciences, University of Lisbon, Portugal}\\
\normalsize{$^{5}$Department of Ecology \& Evolution, University of Lausanne, Switzerland}\\
\normalsize{$^{6}$School of Biological Sciences, University of Southampton, UK }\\
\\
\normalsize{$^\ast$To whom correspondence should be addressed; E-mail:  dap1e12@soton.ac.uk}
}

\date{}

\begin{document} 



\maketitle


\begin{sciabstract}
\textbf{Background.}
The structure and organisation of ecological interactions within an ecosystem is modified by the evolution and coevolution of the individual species it contains. 
Understanding how historical conditions have shaped this architecture is vital for understanding system responses to change at scales from the microbial upwards. 
However, in the absence of a group selection process, the collective behaviours and ecosystem functions exhibited by the whole community cannot be organised or adapted in a Darwinian sense. 
A long-standing open question thus persists: Are there alternative organising principles that enable us to understand and predict how the coevolution of the component species creates and maintains complex collective behaviours exhibited by the ecosystem as a whole? \\
\textbf{Results.}
Here we answer this question by incorporating principles from connectionist learning, a previously unrelated discipline already using well-developed theories on how emergent behaviours arise in simple networks. 
Specifically, we show conditions where natural selection on ecological interactions is functionally equivalent to a simple type of connectionist learning, `unsupervised learning', well-known in neural-network models of cognitive systems to produce many non-trivial collective behaviours. 
Accordingly, we find that a community can self-organise in a well-defined and non-trivial sense without selection at the community level; its organisation can be conditioned by past experience in the same sense as connectionist learning models habituate to stimuli.
This conditioning drives the community to form a distributed ecological memory of multiple past states, causing the community to: a) converge to these states from any random initial composition; b) accurately restore historical compositions from small fragments; c) recover a state composition following disturbance; and d) to correctly classify ambiguous initial compositions according to their similarity to learned compositions.  
We examine how the formation of alternative stable states alters the community's response to changing environmental forcing, and we identify conditions under which the ecosystem exhibits hysteresis with potential for catastrophic regime shifts. \\
\textbf{Conclusions.}
This work highlights the potential of connectionist theory to expand our understanding of evo-eco dynamics and collective ecological behaviours. Within this framework we find that, despite not being a Darwinian unit, ecological communities can behave like connectionist learning systems, creating internal conditions that habituate to past environmental conditions and actively recalling those conditions. 				

\end{sciabstract}

\section*{Background}
With ever-increasing anthropogenic pressure on natural systems, it is vital to understand how the ecosystems we depend upon have been conditioned by evolutionary processes 
in historical environments which may have been very different from those they experience in the present day, and how any such conditioning may shape these systems' responses to new pressures. However, as ecosystems are not typically units of selection, we currently lack a framework linking adaptive pressures on individuals and populations to the dynamical properties of the systems they inhabit. In this article we investigate how systems above the Darwinian levels of selection may evolve collective behaviours, and observe a deep homology with emergent properties well understood in connectionist models of machine learning. We use this homology to develop theoretical analysis of emergent properties of natural selection in ecosystems, and explore the implications for community dynamics through Lotka-Volterra simulation.

\subsection*{Connections and collective behaviours in ecosystems}
The structure and organisation of ecological interactions within biological communities causes them to exhibit many complex behaviours that are not straight-forwardly attributable to the summative behaviour of the individuals they contain \cite{levin1998ecosystems,	levin2005self,	anand2010ecological,	whitham2006framework, ulanowicz1986growth}. For example, the structure of the network of interactions in an ecosystem \cite{neill1974community, 	proulx2005network} affects many of the system's dynamical behaviours including succession dynamics and community assembly rules \cite{clements1916plant, weiher2001ecological},	the stability, resilience and adaptive capacity of a community \cite{holling1973resilience, 	gallopin2006linkages, 	de1995energetics, 	folke2006resilience, 	staniczenko2010structural},  the presence of alternative stable states \cite{higgins2002dynamics, cropp2002ecosystem,beisner2003alternative}, and the system's susceptibility to regime shifts \cite{scheffer2001catastrophic}. 

From some points of view these system-level behaviours exhibit the appearance of design and/or characteristics in common with organismic functions such as development and complex phenotypes \cite{lenton2002gaia, 	ulanowicz1990aristotelean, lenton2004clarifying, richardson1980organismic, mcintosh1995ha}. However, an ecological community is not, in most cases, an evolutionary unit \cite{whitham2006framework, 	mcintosh1995ha, 	levin2011evolution}; it is an assemblage of species each individually adapted to their biotic and abiotic conditions.  Thus the complexity that an ecosystem exhibits is not the product of Darwinian adaptation at the community level \cite{smith1985developmental}. Furthermore, at present we lack general organisational principles that can help us understand and predict how system-level organisation and function results from the many individualistic adaptations on which they depend 
\cite{ anand2010ecological, 	lenton2002gaia, 	mcintosh1995ha, 	gleason1926individualistic, 	lawton1999there, 	milne1998motivation, thompson2001frontiers, 		matthews2011toward},  in particular, the reciprocity between the ecological dynamics on the network and the evolutionary changes to the nodes, and hence, connections of the network 
\cite{ post2009eco,	matthews2011toward,	gross2009adaptive,	paperin2011dual,	schoener2011newest,	turcotte2011impact,	metz1996adaptive,	fussmann2007eco}. 
In short, we do not know how the coevolution of the parts affects the organisation and subsequent behaviour of the whole, i.e. the ecosystem's dynamical properties such as
the location and number of its dynamical attractors;the trajectories it takes towards its these attractors (assembly rules); its stability during assembly and/or succession; and  its sensitivity to initial conditions during assembly. 

Characterising how evolution and coevolution of the parts affects community-level properties is vital to understanding the responses of ecological communities to changes in environment at all scales. This issue is particularly acute in microbial community research, including medical applications in gut flora, where rapid evolution \cite{harcombe2010novel} has the potential to alter the function of those communities we depend upon most intimately, and where there is significant interest in how parental effects create a footprint of community composition that may be remembered throughout life \cite{romano2014maternal}. Coevolutionary processes in gut microbiota have shaped at least three alternative stable states (termed enterotypes) \cite{faust2012microbial}, but it remains unclear how the historical conditioning of different communities' networks of interactions evolve in response to environmental changes in cases such as the use of antibiotics \cite{forslund2013country} and societal changes in diet \cite{david2014diet}, or how these changes affect the emergent properties of community networks \cite{shade2012fundamentals} given the alternate enterotypes that act as attractors for these systems. At the macroscopic scale, Case \textit{et al}. speculate that co-evolutionary processes maintain the distinct bird assemblages on the islands of Bali and Lombok, either side of Wallace’s line \cite{case2005community, mayr1944wallace, wallace2011geographical}. Although birds are relatively unimpeded by the short stretch of sea that has separated terrestrial species, each island maintains distinct avian communities, and the conjecture is that long periods of coevolution within each community has created `coevolved' biogeographic provinces, each network maintains a stable state resistant to invasion by members of the other \cite {case2005community}. Yet, without a framework linking  microevolutionary changes in interactions between species pairs to dynamical behaviours of whole communities,  it remains moot as to whether a network of coevolved interactions could be the explanation for the observed dynamical stability. 

The need to characterise the evolutionary and historical determinants of ecological processes is identified as an important frontier in ecological research \cite{thompson2001frontiers}. Understanding the evolution and adaptability of ecological interactions is necessary, for example, to characterise the response of an ecosystem to climate change or other perturbations \cite{higgins2002dynamics,	staniczenko2010structural,	angeler2010identifying,	carpenter1999management}  and, more generally, to understand how the number and location of dynamical attractors (alternative stable states) are affected by the organisation of ecological interactions acquired over evolutionary time \cite{beisner2003alternative}. 

These issues connect deeply with the phenomenon of \textit{ecological memory} \cite{thompson2001frontiers,	hendry1995role, 	peterson2002contagious,	golinski2008effects} defined by Thompson \textit{et al} \cite{thompson2001frontiers} as ``the result of past environmental conditions and subsequent selection on populations [which] is encoded in the current structure of biological communities and reflected in the genetic structure of species". As an illustrative example, consider the phenomenon of character displacement, \cite{brown1956character,dayan2005ecological}  in which niche divergence between pairs of isomorphic competitors leads to (genetic) trait divergence and increased likelihood of future coexistance; subsequent communities' distributions are shaped by the ``memory" of past competition.
Within ecosystems these genetic memories, distributed amongst multiple populations, influence assembly processes, stability, and resilience of the developing and mature ecosystem \cite{thompson2001frontiers}.  However, there remain many open questions about exactly how the microevolutionary modification of interspecific relationships by natural selection shapes any macroevolutionary memory at the community level 	\cite{thompson2005geographic, thompson2001frontiers}. For example: 
\begin{enumerate}
\item	How do changes to interactions evolved in past environmental conditions alter the response of the community to future changes or perturbations in environmental conditions? 
\item	Is ecological memory merely a passive memory (like an imprint in clay) where the persistent effects of the most recent ecological states over-write or blend with those of older states, or can an ecological memory retain information about multiple distinct past states without just averaging them? 
\item Can the assembly rules and succession dynamics of a community be systematically organised by selection in past environmental states? 
\item How does the formation of an ecological memory affect the possibility of alternative ecological stable states, and regime shifts under subsequent environmental forcing?
\end{enumerate}
The lack of a theoretical framework that links individual adaptations to collective behaviours leaves such questions unanswered. Our aim in this paper is thus to introduce such a framework. We do this by converting and exploiting theory that is already well-developed in another domain, namely, connectionist models of memory and learning. Below, we discuss the characteristics of connectionist models and their relationship to eco-evo dynamics. We then show a formal equivalence between these systems. We conclude that community-level organisation does not require community-level selection. The organisation of a community can be conditioned by past experience (collectively habituated to past environmental conditions) in the same sense, and with the same consequences for collective behaviours at the community level, as connectionist models of memory and learning. In order to demonstrate how making this general link between these disciplines leads to new insights about specific ecological behaviours, we then simulate mathematical models derived from this framework to address the above open questions about ecological memory.

\subsection*{Connectionist models of memory and learning }
\textit{Connectionism} is an approach to modelling cognition, in particular using neural networks, that explains how complex system-level behaviours can arise via the appropriate organisation of many simple components. The first important contribution of these models is to show that although each unit in a network might be very simple (e.g. the activation level of a neuron is simply a non-linear sum of the weighted connections from other neurons \cite{mcculloch1943logical,	hinton1999unsupervised,	hopfield1982neural}), if appropriately organised/connected,a network of such units can provide many remarkable collective behaviours, including: a) forming a distributed memory for one or more configurations; b) pattern recognition from partial stimulus; c) the removal of noise from corrupted compositions; and d) classification of ambiguous inputs \cite{mcculloch1943logical,	hinton1999unsupervised, 	hopfield1982neural, watson2010optimisation}. 	It has been noted in many different domains that the collective behaviours that can be exhibited by neural networks are not exclusive to neural models and can be exhibited by other types of dynamical systems (e.g. gene regulation networks, immune systems, multi-agent systems, economic systems and social networks) 
\cite{farmer1990rosetta, 	watson2011global,	mikhailov1990molecular,	fernando2009molecular,	vohradsky2001neural,		noonburg1989neural,	poderoso2007model}. This includes ecological networks (where the growth rate of a species is modelled as a non-linear sum of the weighted fitness-interactions with other species) 	\cite{wilson1992complex,	noonburg1989neural,	poderoso2007model}.
	
A deficit in the analogy between neural networks and ecosystems is that whereas neural networks acquire the organisation necessary for their collective behaviours through learning mechanisms designed for that purpose, ecological connections are modified by individual-level natural selection with no such system-level purposes in mind. Although ecological networks may have population dynamical similarities with neural activation dynamics in neural networks, there has not been any reason to expect that both systems may be organised in a similar manner. However, connectionist models also show that network organisations sufficient for many collective behaviours can be generated via learning mechanisms that modify the strength of connections according to only very simple and local reinforcement principles – even by mechanisms that do not require any system-level reward or performance-based feedback. The full significance of this for the evolution of ecological networks has not been previously appreciated \cite{watson2014evolution}.

Learning mechanisms in neural network models have two basic types \cite{watson2014evolution}. \textit{Supervised} learning utilises an external reward signal, or error function, to direct incremental changes to connections. We have recently demonstrated a formal equivalence between supervised learning and the evolution of connections in a network that is selected (at the system level) to produce a particular target phenotype or phenotypes \cite{watson2011global,	watson2014evolution}. However, in the absence of a group selection mechanism there is no “target” phenotype directing selection at lower levels within ecological communities; supervised learning does not occur at this scale. 

The other type of learning in these systems is \textit{unsupervised} learning (Box 1), which operates without a reward signal. This may seem counter-intuitive but, when learning correlations or associations, learning what things ‘\textit{often} go together’ has many useful properties that can be attained without a supervisory signal to indicate what things ‘\textit{should} go together’ \cite{hinton1999unsupervised,	hopfield1982neural,	watsoninpreplearning,	ackley1985learning}.	 Thus, whereas supervised correlation learning reinforces correlations that are \textit{good} according to some external reward signal, unsupervised correlation learning changes connections simply to reinforce correlations that are \textit{frequent}. Hebbian learning \cite{hebb1949organization} 	is the simplest unsupervised correlation learning mechanism and is well-understood in neural network models of memory and knowledge representation \cite{hinton1999unsupervised,	hopfield1982neural,	ackley1985learning}. 	Under Hebbian learning, the change in strength of a synaptic connection, $\Delta \omega_{ij}$, is proportional to the co-activation of the neurons it connects: i.e.  $\Delta \omega_{ij} = r x_i x_j$, where $r> 0$ is a learning rate, and $x_k$ is the activation level of node $k$. This type of learning instantiates a very simple positive feedback principle between behaviour and connections, often paraphrased as ``neurons that fire together wire together". The effect of such changes is that \textit{correlation becomes causation}, i.e. variables that happen to be both active at the same time (e.g., because they are stimulated by the same external conditions) become causally related by connections internal to the system, and thus their behaviour becomes more correlated in future. In this manner the network habituates to the perturbations it experiences by internalising information about the pattern of perturbation it has experienced into the organisation of its connections.

\begin{figure}[p!]
\begin{mdframed}[skipabove=0pt,leftmargin=32pt]
\footnotesize
\textbf{Box 1: Hopfield networks and unsupervised learning using Hebb's rule.}\\
The Hopfield network model \cite{hopfield1982neural} originated from the hypothesis that it is the structure of connectivity between units in the central nervous system, rather than differences between the units themselves, that is most important in understanding the brain's complex behaviours. 
These simple models are fully connected networks of identical units. As units are identical it is solely differences in connections between units that determines the behaviour of each network. 
Despite this simplicity, these systems display complex behaviours, including the capacity to form multiple distributed memories;  indeed, they are the simplest systems that have this capacity. 
Hopfield networks (and neural networks in general) are able to form multiple memories of configurations because,
 for each memory they store \textit{correlations} between units, rather than the \textit{states} (or outputs) of units. 
An effect of this architecture is that the structure of a Hopfield network can be updated to enable it to learn new patterns without over-writing and destroying pre-existing memories (figure \ref{fig: hopfield_box}).

Hebb's rule is an unsupervised learning technique that can be used to train Hopfield networks to form memories for one or more configurations. 
For each training pattern Hebbian learning alters connections between units in the direction that reinforces the correlations between those unit's current outputs. 
If two units have the same state in multiple patterns they will become strongly correlated; Hebb's rule causes units that `fire together' to become `wired together'.
Hebb's rule is an `unsupervised' process because it does not utilise quality functions on the data used to train a network (whereas a supervised process might, for example, use a quality metric to scale changes made to the network).
That is, Hebbian learning only acts to reinforce `frequent' correlations in the training data, rather than correlations that are `optimal' according to some metric of system performance (as in supervised learning methods). \\

\begin{center}
\includegraphics[width = 0.8\textwidth]{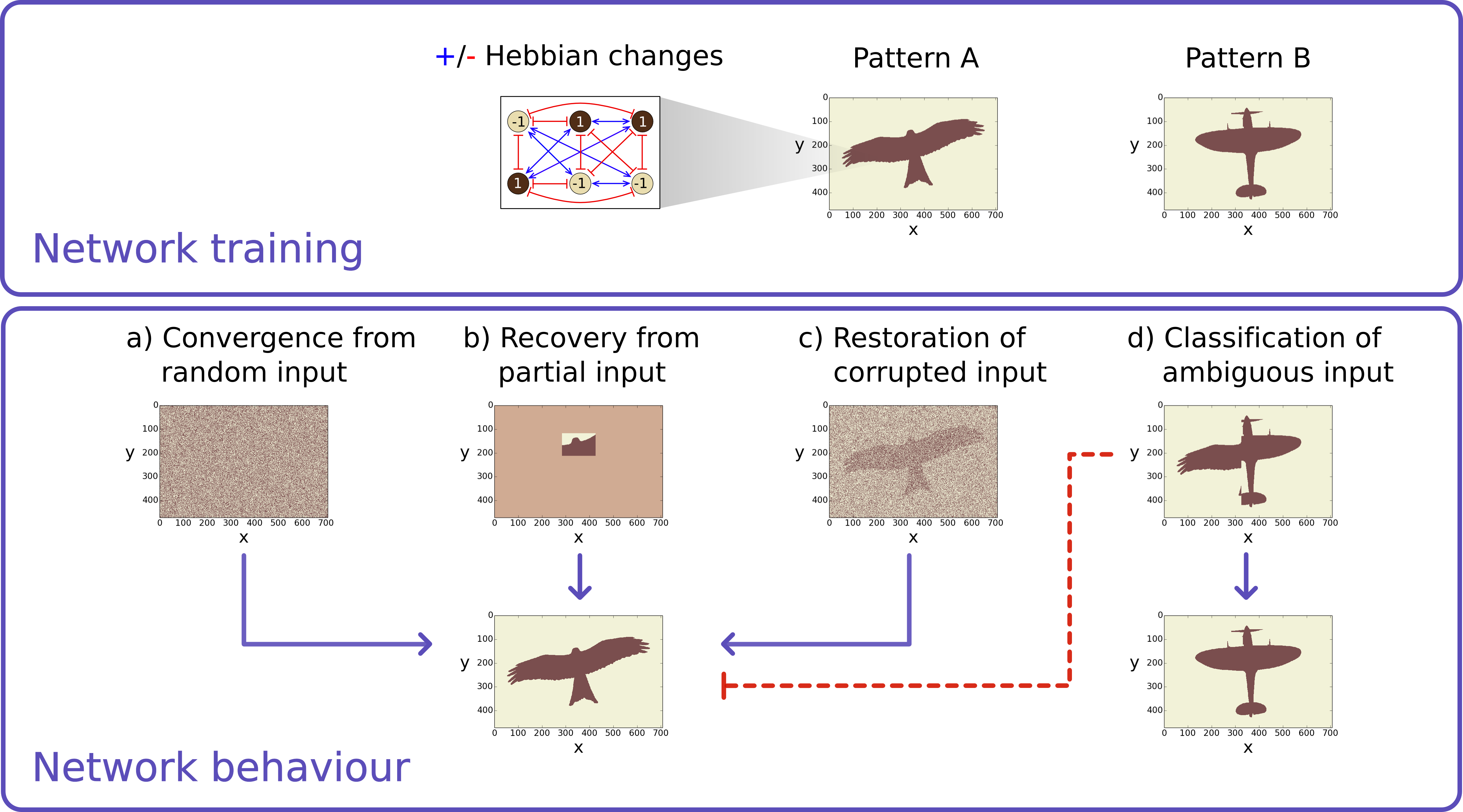}
\end{center}
  \caption{\footnotesize
  \textbf{Network training}: Unsupervised learning processes as used to train a Hopfield network to store two configurations, patterns A and B. 
  Each unit in the Hopfield network corresponds to a pixel in the image display.
  Six units are highlighted to illustrate the changes to connections during training in pattern A. 
  Hebbs rule alters connections between units such that units of the same sign (1:1 or -1:-1) become more correlated (blue lines) and units of opposite signs (1:-1 or -1:1) become more anti-correlated (red lines).
  \textbf{Network behaviour}: Training the network on both patterns results in a network with attractors (a.k.a. memories) for these patterns and system dynamics result in all initial conditions converging to one of the trained patterns (a).
  This behaviour enables these systems to be used for a variety of functions, including: 
  b) recovery of complete composition from partial input; 
  c) noise reduction; and 
  d) classification (the input image is a closer match for the plane configuration than the bird configuration).}
  \label{fig: hopfield_box}

\end{mdframed}
\end{figure}

This simple principle is capable of producing many remarkable collective behaviours elucidated over more than 50 years of neural network research \cite{hinton1999unsupervised,
	hopfield1982neural,	ackley1985learning}. 	Famously, this includes the ability to develop of a \textit{distributed associative memory} which can store and recall multiple patterns of activation in the organisation of synaptic connections \cite{hopfield1982neural}, facilitating the use of these networks use in pattern recognition, noise reduction and classification (fig \ref{fig: hopfield_box}). 	A main contribution of this paper is to show that in ecological communities, given heritable variation in ecological relationships and certain conditions on ecological constraints, these positive feedback principles obtain from the action of individual natural selection \cite{ulanowicz1995utricularia}. Table 1 sets out the full analogy we make between connectionist learning in neural networks and eco-evo dynamics in ecosystems, starting with the previously recognised dynamical equivalence (Table 1.a-f). 	
\\

\linespread{1}\selectfont

\begin{table}[p!]
\scriptsize
\caption{Mechanistic equivalence between evo-eco dynamics and learning neural networks, and a map for the comparisons and analogies made in this paper. 
	a-f) The basic components of the analogy made in the introduction to this paper. 
	g) The main contribution of this paper (discussed in part I) – the equivalence of individual natural selection acting on inter-species interactions with a simple associative learning rule such as Hebbian learning. Thus ecological networks \textit{evolve} like neural networks \textit{learn} (fig. \ref{fig: Hebbian equivalence}). 
	h-m) From this the phenomenology shown in our experiments follows (simulation results, fig. \ref{fig: experiment 1} \& \ref{fig: Experiment 2}).
}
	
	\setlength{\extrarowheight}{0.1cm}
    \begin{tabular}{p{.05\textwidth}p{.45\textwidth}p{.45\textwidth}}

&    \normalsize Unsupervised correlation learning 
&  \normalsize Coevolution  \\ 
\hline 
\hline
\rowcolor{blue!10} &     Activation dynamics & Population dynamics \\ \hline
a)&    Neural activation level                                                                                                                                                                                                        
& Species density, $x_i$
\\
\hline
b) 
&    Neural activation pattern                                                                                                                                                                                                      
& Ecological state, $X= \{ x_i, x_2, ... x_N \}$                                                                                                                                                                                                                                                              \\
\hline
c) 
&    Synaptic connection strength, $\omega_{ij}$                                                                                                                                                                                                 
& Inter-species fitness interaction, $\omega_{ij} $                                                                                                                                                                                                                                           \\
\hline
d)
& Neural network \newline (weight matrix, W).                                                                                                                                                                                             
& Ecological network \newline (community matrix, $\Omega$).                                                                                                                                                                                                                                        \\
\hline
e)
& Neural activation dynamics: a non-linear weighted sum of inputs from other neurons (and external inputs).                                                                                                                         
& Ecological population dynamics (Eq. 1): species growth is a non-linear function of the sum of weighted fitness interactions from other species (and environmental changes to carrying capacities).                                                                               
\\
\hline
f)
& External input patterns \newline (aka. `training set').                                                                                                                                                                              
& Environmental forcing \newline (in multiple environmental conditions)                                                                                                                                                                                                                      \\
\hline
\hline
\rowcolor{blue!10}
&    Correlation learning \newline (unsupervised)                                                                                                                                                                                              
& Evolution of interactions \newline (individual selection only)                                                                                                                                                                                                                             \\
\hline 
g)
&
    Positive feedback between activation strengths and connection strengths – aka. \textit{neurons that fire together wire together}. Unsupervised correlation learning mechanism, Hebb's rule: $\Delta \omega_{ij} = r x_i x_j$, where $r > 0$ is a learning rate.  
& Positive feedback between ecological densities and connections – or \textit{species that occur together wire together}. Direct effects of individual natural selection on interactions: $v_{ij} =  r x_i x_j$, where $r =  \frac{m_i}{k_{ie}} g \mu$  describes the available mutation (Eq. 3).  
\\
\hline \hline
\rowcolor{blue!10}
&    Collective behaviours in neural networks  \newline (arising from e.g., Hebbian learning, Fig \ref{fig: hopfield_box})
& Collective behaviours in ecosystems  \newline (arising from individual selection acting upon interspecific correlations)                                                                                                                                                                                                                                               \\
\hline
h) 
& \textit{Memory formation} (fig \ref{fig: hopfield_box}, top panel) Hebb's rule organises synaptic connections to reinforce the state of the system, decreasing sensitivity to changes in input.
&
\textit{Ecological memory formation }(fig 3): natural selection organises ecological relationships in a manner that reinforces the current ecological state, decreasing sensitivity to changes in environmental conditions. (Attractors due to environmental variables become attractors of community dynamics \cite{beisner2003alternative}.)
\\
\hline
i) 
&    \textit{Distributed associative memory facilitates a memory of multiple patterns} (fig \ref{fig: hopfield_box}.a): the capacity to store multiple patterns of activation in the organisation of synaptic connections and recall patterns from any initial conditions via activation dynamics.
& 
\textit{Formation of alternative stable states} (fig. \ref{fig: Experiment 2}.a): the creation of a distributed ecological memory in the network of species interactions results in a system with attractors that mimic past ecological states. 
\\
\hline
j) 
&   \textit{Pattern reconstruction} (fig \ref{fig: hopfield_box}.b): the recall of a complete pattern from a partial stimulus.
& \textit{Ecological assembly dynamics } (fig. \ref{fig: Experiment 2}.b): reconstruction of a particular community composition, from a subset of that community.\\
\hline
 k) 
 &\textit{Error correction} (fig \ref{fig: hopfield_box}.c): the ability to remove noise from a pattern, repair imperfections and restore a complete pattern. 
 & \textit{Ecological resilience} (fig. \ref{fig: Experiment 2}.c): the ability to recover from perturbations in species densities and restore the complete community. 
  \\
\hline
l)
&   \textit{Recognition} or classification of an input or stimulus (fig \ref{fig: hopfield_box}.d): return the nearest attractor from ambiguous initial conditions. 
& \textit{Ecological sensitivity} to initial conditions  (fig. \ref{fig: Experiment 2}.d): the switch-like change in response to small variation in initial species densities.  \\
\hline
m) & \textit{Holding state in dynamics}: Hopfield networks and other recurrent networks have an internal state that allows them to display temporal dynamics (independent of input). 
& \textit{Ecosystems hold state in population dynamics} (fig. \ref{fig: Hysteresis}): in systems with multiple attractors this results in a communities capable of hysteresis with tipping points between states.  
\\
    \end{tabular}
\end{table}

The biological evidence for positive feedback between ecological and evolutionary dynamics is entirely intuitive and already recognised in many areas where populations shape their future selective pressures (e.g. niche construction \cite[and refs. within]{laland2014role}) but the full implications of this feedback have not been realised \cite{laland2009conceptual}. For example, this feedback is part of the backstory involved in ‘invasional meltdown’ \cite{gallardo2015great,	heimpel2010european,	simberloff1999positive}	where species that have been in prolonged contact with one another in one environment facilitate one-another's invasion into a new environment because they ``have had a long evolutionary time to develop a cosy relationship with each other" \cite{gallardo2015great}. 	Notice the simple positive feedback involved; species that occur in high density at the same time and under the same environmental conditions coevolve to become less competitive with each other over time. In turn, this reduction in niche overlap makes it more likely that they will coexist in high-density together in future. Our first key result is to formalise this principle with population genetics and show its equivalence with unsupervised correlation learning (Table 1.g). Then, to demonstrate how this opens-up a transfer of concepts and results between these domains, we use numerical simulation to show conditions where an ecosystem can acquire, hold and recall distributed information about past environmental conditions – i.e. form a distributed ecological memory. This demonstrates several phenomena that are well-understood in connectionist models of memory and learning (Table 1.h-m).

The formal link between the disciplines does not depend on the specific scenarios relevant to investigating ecological memory nor on the simplifications that are necessary for the simulation models; for this reason we divide our work into two parts: I) an analytic model and results concerning the general equivalence, II) simulation methods and results concerning ecological memory.

\section*{Methods Part I: A Model}
\subsection*{Ecological dynamics}
We model an ecosystem state as a vector of population densities over all $N$ species, $X=\{x_1,x_2,… x_N\}$, ($x_i \ge 0$), and an interaction network, or ``community matrix" \cite{neill1974community}, $\Omega$, where each element, $\omega_{ij}$, represents the fitness effect of species $j$ on species $i$ relative to $i$ on itself ($\omega_{ii}$ = -1). We assume $\forall i, j :\omega_{ij} \le 0$, e.g. competitive (non-trophic) interactions such as via competition for resources. A Lotka-Volterra competition equation (Eq. 1) defines the rate of change of density of a species as a function of its intrinsic growth rate and a weighted sum of interactions with all other species \cite{wilson1992complex}:
\begin{equation}
\frac{dx_i}{dt} = \frac{m_{i} x_i}{k_{ie}} + \left( k_{ie}  + \sum_{j=1}^{N} \omega_{ij} x_j  \right)
\end{equation}
where $x_i$ is the density of species $i$, $m_i$ is the intrinsic net growth rate of species $i$, $k_{ie}$ is the carrying capacity of species $i$ in environment $e$ (i.e. its density before interspecific competition), and $N$ is the number of species in the network. The dynamical equivalence between models of this sort, where populations experience exponential growth asymptotically approaching a threshold, and those models used in neural networks of excitation/inhibition between neurons is well recognised \cite[and refs. within]{noonburg1989neural}. 
We now turn our attention to the selective pressures on interspecific interactions, and make new comparisons with unsupervised learning in neural networks. 

\subsection*{Evolution of interactions under individual selection}
Each interaction coefficient summarises how a variety of structural, physiological and behavioural traits affect the degree to which one species impacts the population growth of another. Although subject to bio-physical constraints, e.g., stoichiometric constraints on resources, these interactions can often be modified by the evolved characteristics of the constituent species, e.g. traits that alter the overlap of habitat preference or resource utilisation profiles \cite{hutchinson1965niche}	or the time, effort or energy expended on a particular ecological resource or relationship. 
	
We assume that only individual-level selection acts on these interactions. We do not model selection on whole ecosystems (e.g., via a population of ecosystems), nor on species. Thus only changes to traits that directly affect the growth rate of an individual compared to the rest of the individuals in the species can be selected. Individual selection acts to decrease the competitive effects from others by changing $\omega_{ij}$; but note that an individual has no intrinsic interest in altering the growth rate of others by changing $\omega_{ji}$. Changes that decrease the density of a competitor, for example, cannot be selected for under individual selection as (in the absence of group selection) such changes benefit all individuals in a species \cite{wilson1980natural}. Therefore any changes to a species growth rate that occur as a side-effect of altering the density of some other species (e.g., via changes to $\omega_{ji}$ or via ecological trade-offs below) are not affected by individual selection (Appendix a).  

\subsection*{Analysis of individual-level natural selection acting on ecological interactions}

We analyse the rate of accumulation of favourable mutations, $v$, in each interaction coefficient, $\omega_{ij}$. In order to study the dynamical interaction between evolutionary and ecological dynamics, we are particularly interested in how the evolution of $\omega_{ij}$ is sensitive to the current species densities. The qualitative picture is as follows: Occasionally, mutants arise in species $i$ that are identical to $i$ except for the modification of an interaction coefficient with another species $j'$ in the ecosystem. The origin and establishment of such a mutant can be modelled by applying population genetics theory \cite{neher2010rate, 	weissman2012limits}	to the particular case. From the ecological dynamics it follows that the selective coefficient, $s$, conferred by the change, $g$, in the interaction coefficient $\omega_{ij'}$, is the change in the invasion rate per capita of a mutant type of species $i$ relative to the growth rate per capita of species $i$ without the mutation:
\begin{equation}
s=\frac{m_{i}}{k_{ie}} \left( 
k_{ie} + \sum_{j=1}^{N}  \omega_{ij} x_j + g x_{j'}\right) -
\frac{m_{i}}{k_{ie}} \left( 
k_{ie} + \sum_{j=1}^{N}  \omega_{ij} x_j \right)= 
\frac{m_{i}}{k_{ie}} g x_{j'}
\end{equation}
(Simplified as $s = \frac{m_i}{k_{ie}}g x_{j}$ henceforth.) Since $m$, $k$ and $x$ are positive, a favourable mutation requires only $g> 0$. Qualitatively, this means that a mutation to an individual of one species, e.g. a change in its habitat or resource usage, is selected for if the mutation reduces the negative influence of another species on its growth rate. We assume that in all species such mutations occur at rate $\mu$ per individual per generation.  In general, the rate of accumulation of such mutations is equal to the product of the number of individuals, $x_i$, the beneficial mutation rate, $\mu$, and the average probability that a single new mutation will ultimately fix, $\bar{P}$, such that: $v = x_i \mu \bar{P}$ \cite{	weissman2012limits}. 	In large sexual populations with linked loci, $\bar{P}$ will depend on $v$, and in different ways depending on the type of recombination, recombination rate, population size, the mutation rate and magnitude of mutations 	\cite{neher2010rate, 	weissman2012limits}.	For unlinked loci, in small populations, or under strong selection and weak mutation where mutations occur serially, $\bar{P}$ is proportional to the selection coefficient, $s$ 	\cite{	weissman2012limits}. Since the effects we want to investigate do not depend on the effects of sexual recombination it is sufficient for our purposes to model the rate ofadaptation in this simple manner. In this case, the rate of adaptation, $v_{ij}$, in an interaction coefficient, $\omega_{ij}$, is given by:
\begin{equation}
v_{ij} = x_i \mu s = \frac{m_i}{k_{ie}} g \mu x_i x_j
\end{equation}
In more complex cases, where there is interference between alleles at different loci, $v_{ij}$ may not be linearly proportional to $x_ix_j$ as it is in Eq.3, but in all cases, the rate of evolutionary change in an interaction coefficient increases with the product of $x_i$ and $x_j$ since mutations must be both created and selected in order for an interaction coefficient to evolve.  This is robust to the choice of underlying model (Appendix b). This is entirely intuitive: a) if suitable heritable variation in relationships is available, natural selection always acts to reduce the negative effects of others, and b) the rate of adaptation of the interaction coefficient between two species, e.g. by character displacement, is driven by their co-occurrence \cite{thompson2005geographic}. 

This is our first key result, describing how selection acts on inter-species relationships as a function of the current ecological state (Table 1.g).  Eq. 3 tells us that the rate of adaptation on inter-species relationships is proportional to the co-occurrence of the species involved: Hence, \textit{species that occur together} (arise in high density at the same time and under the same conditions), \textit{‘wire’ together} (and there will be selection for changes to interactions that makes those species more likely to co-occur in future) - as per the principle of unsupervised correlation learning. (Correlation learning can be produced either by a \textit{reduction} in negative interactions, as here, or by an \textit{increase} in positive interactions, with the same effect on system dynamics, i.e. either will increase the future co-occurrence of the species that have co-occurred in the past.) We now investigate the consequences of this finding for collective behaviours in an ecological community, using ecological memory as a case study.

\section*{Methods Part II: Simulation}

In general, Eq.1 may exhibit unstable or even chaotic behaviour. In the following investigations we restrict our simulations to interactions that are symmetric ($\forall i,j : \omega_{ij} = \omega_{ji}$) as per competition for shared resources or for competition coefficients estimated from utilisation functions \cite{may1975some}, in which case the dynamics have only fixed point equilibria \cite{hughes1998aggregate}. During simulation we allow the ecological dynamics to equilibrate at each time step (over $\tau$ iterations of Eq.1). Then all interaction coefficients are updated according to the direct effect of natural selection in proportion to the rate of adaptation (Eq. 3). Then ecological constraints are applied to these interactions as follows, and the process is repeated.

\subsection*{Ecological constraints/evolutionary trade-offs on changes to ecological interactions}
In ecosystems where niche space is saturated, the capacity of natural selection to alter interactions is subject to inevitable ecological constraints and evolutionary trade-offs that prevent selection from eliminating all competition. Individuals with traits that cause them to avoid competition with one species  may be forced to compete more with others. Thus the interaction between two species is more generally governed by a) the evolvable characteristics of the species as described by Eq.3, and b) evolutionary trade-offs or ecological constraints applied by the physical properties of the environment (e.g., energy spent on exploiting one resource cannot also be spent exploiting another). Here these trade-offs are represented by normalisation conditions that conserve the sum of interactions to and, by symmetry, from each species. Specifically, for all species $i$ and $j$ ($j \ne i$), $\sum_{j=1}^N \omega_{ij}(t) = Q_i$ and $\sum_{j=1}^N \omega_{ji}(t) = Q_j$, where $Q_i = Q_j < 0$ is a constant (the sum of interaction terms in row/column $i$ at time $t=0$) (see Appendix c). Such normalisation represents ecological niches that resist change in width more than change in location, e.g. individuals can more easily change which resources they depend on than how dependent they are overall \cite{roughgarden1972evolution}.
	 
Although natural selection always acts to reduce competitive impacts from others, the fact that the rate of adaptation is greater for some competitive interactions than others (Eq. 3), together with these normalising evolutionary trade-offs, will mean that the competition between some species will increase. When the interaction, $\omega_{ij}$, from some species $j$ to a given species $i$ is, for example, made \textit{less} competitive (decreased in magnitude) by the evolution of heritable traits, all the other interactions involving $i$, i.e., $\omega_{ih}$ ($h \ne j$) and $\omega_{hi}$ ($h \ne i$), are made more competitive by these normalising evolutionary trade-offs.  This, in turn, leaves all interactions not involving $i$ relatively less competitive.  Self-interactions are not modified by either evolutionary or normalisation mechanisms ($\omega{ii}=-1$).

\subsection*{Environmental forcing}
  \begin{figure}[h!]
  \begin{center}
  \includegraphics[width = 0.8\textwidth]{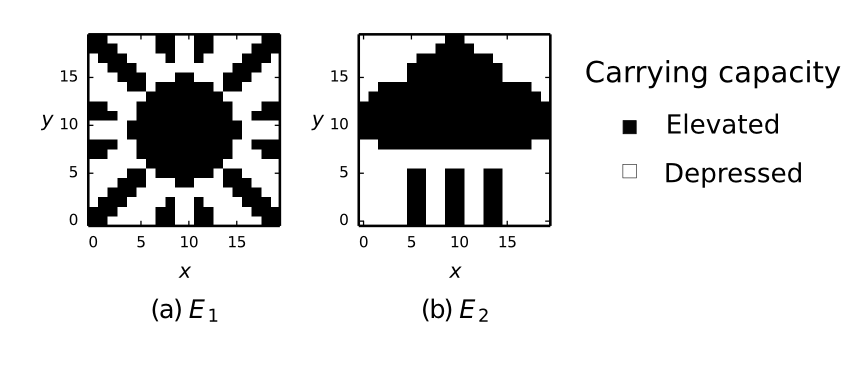}
  \end{center}
  \caption{\textbf{The carrying capacities of 400 species are affected differently by two different environmental conditions, $E_1$ and $E_2$.}
      For our purposes, the specific patterns of carrying capacities for the two conditions are arbitrary and can thus be depicted by $20 \times 20$ pixel `hot' and `cold' pictograms where a black pixel at location $x,y$ indicates an elevated carrying capacity, ($k_0 + \alpha$), and a white pixel a depressed carrying capacity, ($k_0 -\alpha$), for the ($20x+y+1)^{th}$ species in the ecosystem, Appendix c (the two-dimensions of the pictogram are abstract and do not imply any spatial arrangement of the species).
      }\label{fig: pictograms}
      \end{figure}

To investigate ecological memory we are interested in how the evolution of ecological interactions is influenced by past ecological states. To model the evolution of an ecosystem under varying environmental conditions that force or drive the ecosystem to adopt different ecological states, we define two environmental conditions, $E_1$ and $E_2$, that have differing effects on the carrying capacities of the species (Appendix c). Relative to a default environment $E_0$, environment $E_1$ increases the carrying capacity of some species and decreases others, whilst in $E_2$, a different subset of species is increased/decreased. $E_1$ and $E_2$ may represent hot-dry and cold-wet climates, for example; or high/low levels of some key broadly-utilised resource such as phosphorous input rates for a lake habitat  \cite{carpenter1999management}. Given that individuals from each species experience both conditions over evolutionary timescales, these conditions could vary in space (e.g. geographic localities, \cite{paperin2011dual}), rather than in time (e.g. seasonal change). To make the effects of these two conditions on community composition easily identifiable we utilise environmental forcing patterns corresponding to two arbitrary but easily identifiable pictograms (Fig. \ref{fig: pictograms}). Here the \textit{hot} and \textit{cold} pictograms describe two different configurations of species densities representing, for example, hot dry savannah and cold wet/temperate ecological states, respectively. The environment is switched between $E_1$ and $E_2$ every $T$ evolutionary updates.\\
\vspace{1cm}

Model parameters of the simulations and methods used for assessing ecological attractors are described in Appendix c.

\section*{Results}
We use the series of four open questions concerning ecological memory listed above to exemplify some of the implications of our general result.

\begin{figure}[hp]
\begin{center}
 \includegraphics[width = 0.9\textwidth]{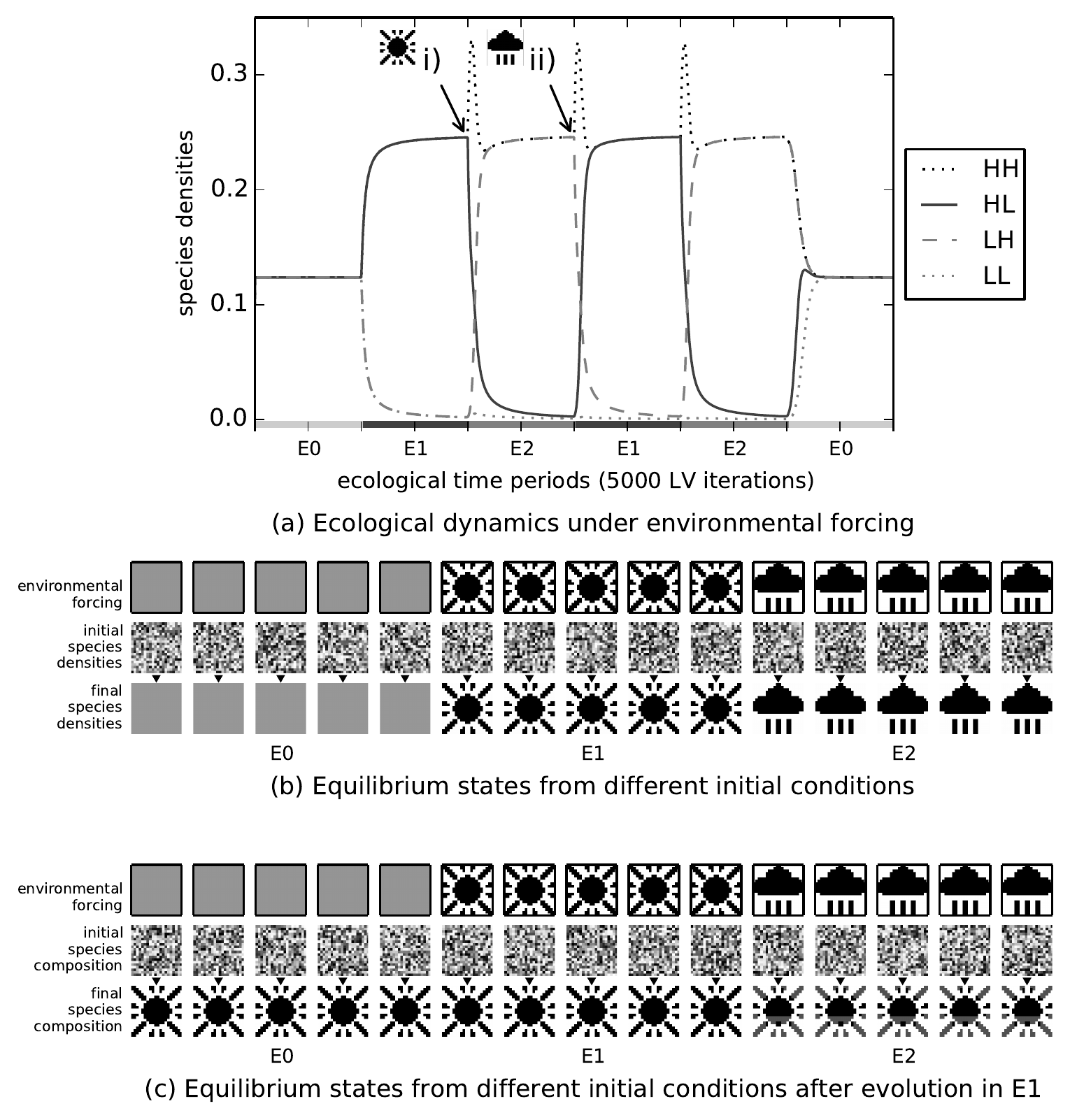}
\end{center}

  \caption{\textbf{Ecological dynamics before and after evolution in $E_1$.}
      a-b) Before evolution of interactions, when forcing is applied, some species densities increase, others decrease. a) Four species responding differently to $E_1$ and $E_2$ (H=`high', L=`low'). 
      b) Vectors of all species population densities are displayed in a pixel array as per Fig. \ref{fig: pictograms}. Under a given pattern of environmental forcing (top row), an initially random pattern of species densities (middle row), equilibrates at a pattern of species densities (after $\tau$ timesteps) (bottom row). Initial species densities do not alter the attractor attained (5 independent examples). 
      c) After evolution of interactions in $E_1$, equilibrium states are governed by that past pattern of environmental forcing and not by the current environment. This ecological memory is a stable attractor, reached from any initial pattern of species densities, regardless of the pattern of environmental forcing (some distortion is visible under $E_2$ forcing).}
       \label{fig: experiment 1}
      \end{figure}

\subsection*{i) Changes to interactions evolved in past ecological states ‘canalise’ the response of the community to subsequent changes in environmental conditions or future perturbations.}
Experiment 1 investigates how evolution in fixed environmental conditions changes the ecological dynamics of the community. Before the evolution of interactions, during the ecological phase of simulation, the ecosystem arrives at a stable equilibrium corresponding to the pattern prescribed by the current environmental forcing (Figs. \ref{fig: experiment 1}.a and \ref{fig: experiment 1}.b). Inter-species interactions are then evolved in environment $E_1$, i.e., without changes to the environmental forcing during evolution.  The process is repeated for 800 ecological and evolutionary cycles. We then assess how evolved interactions have altered the sensitivity of the ecosystem to subsequent environmental forcing. We find that the ecosystem now arrives at a stable equilibrium corresponding to the $E_1$ pattern (the pattern it experienced when interactions were evolving) even when the environmental forcing is subsequently changed to $E_0$ or $E_2$ (Fig. \ref{fig: experiment 1}.c). Experiment 1 thus shows that the effect of evolving ecological interactions by individual natural selection under fixed environmental conditions is to create a stable attractor for the specific ecological state experienced in that past environment, reducing the responsiveness of the ecosystem to respond to subsequent environmental forcing, and increasing the adaptive capacity of the system to withstand changes to environmental conditions or perturbations to population densities.  This behaviour demonstrates the basics of an ecological memory, but only a memory of one pattern. Even passive systems can remember one pattern, e.g. an imprint in clay, but connectionist models show that a dynamical network is capable of storing and recalling multiple patterns.\\

\subsection*{ii) Ecological memory can retain and recall information about multiple distinct past states. }
\begin{figure}[h!]
\begin{center}
 \includegraphics[width = 0.9\textwidth]{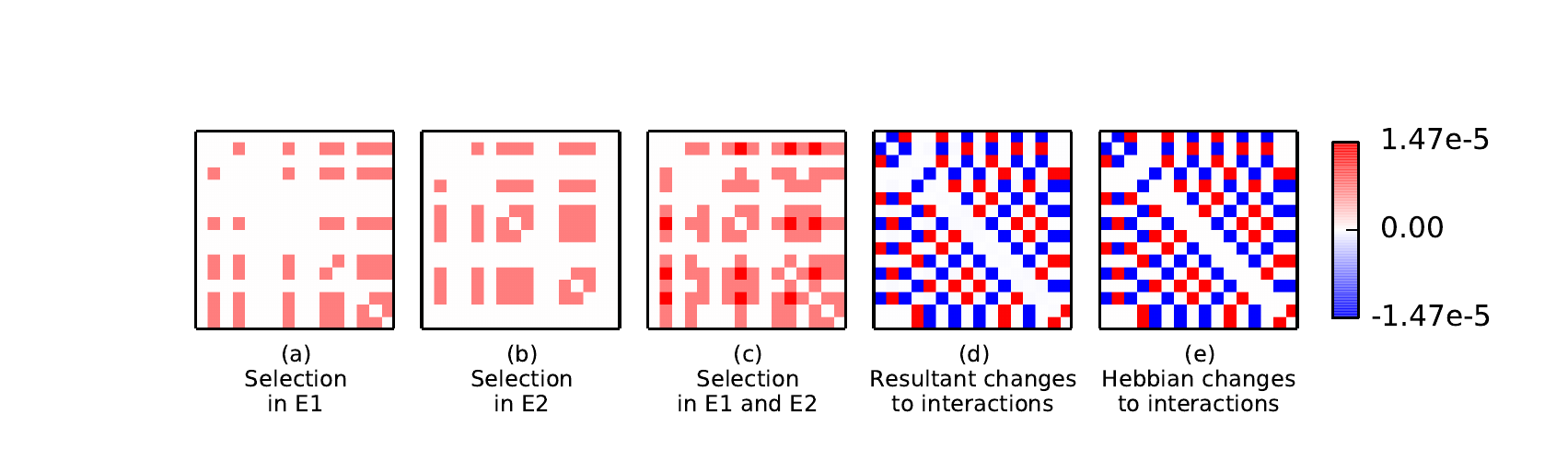}
\end{center}
  \caption{\textbf{Evolved interactions are identical to Hebbian interactions.}
      Change in interactions between the first 16 species are shown. 
      a-b) Some of the competitive interaction coefficients are decreased by the direct effects of selection in $E_1$ and $E_2$, respectively. 
      c) The combined effect of selection in the two environments is that some interactions are decreased in both environments, some in only one environment and others in neither environment. This depicts the relative rate of change due to direct selection effects. 
      d) When normalising ecological constraints are taken into account, some interactions are decreased, some left unchanged, and others are increased. The resulting changes are identical to (e). 
      e) The result of Hebb's rule applied to the interactions between the first 16 species summed over $E_1$ and $E_2$ ($r$ is scaled to give the same mean magnitude as (d)).}
       \label{fig: Hebbian equivalence}
\end{figure}

In Experiment 2 varying environmental conditions are applied to cause the ecosystem to adopt two different ecological states ($E_1$ and $E_2$) repeatedly whilst inter-species interactions are evolving. The effect of these evolved changes plus normalising evolutionary trade-offs are illustrated in Fig \ref{fig: Hebbian equivalence}. We see that their evolution is identical to Hebbian learning (again this is robust to the choice of underlying model, Appendix d).

After evolution we find that, in the absence of further environmental forcing, the ecological dynamics have two stable attractors corresponding to $E_1$ and $E_2$, reached from any initial species densities (Fig. \ref{fig: Experiment 2}.a., Appendix c). An ecological memory can thus retain information about multiple distinct past states without just averaging them or blending them (for example, the system does not have an attractor for the union of both patterns).


\subsection*{iii) The assembly rules of a community can self-organise to recreate past environmental states.}

\begin{figure}[hp]
\begin{center}
 \includegraphics[width = .9\textwidth]{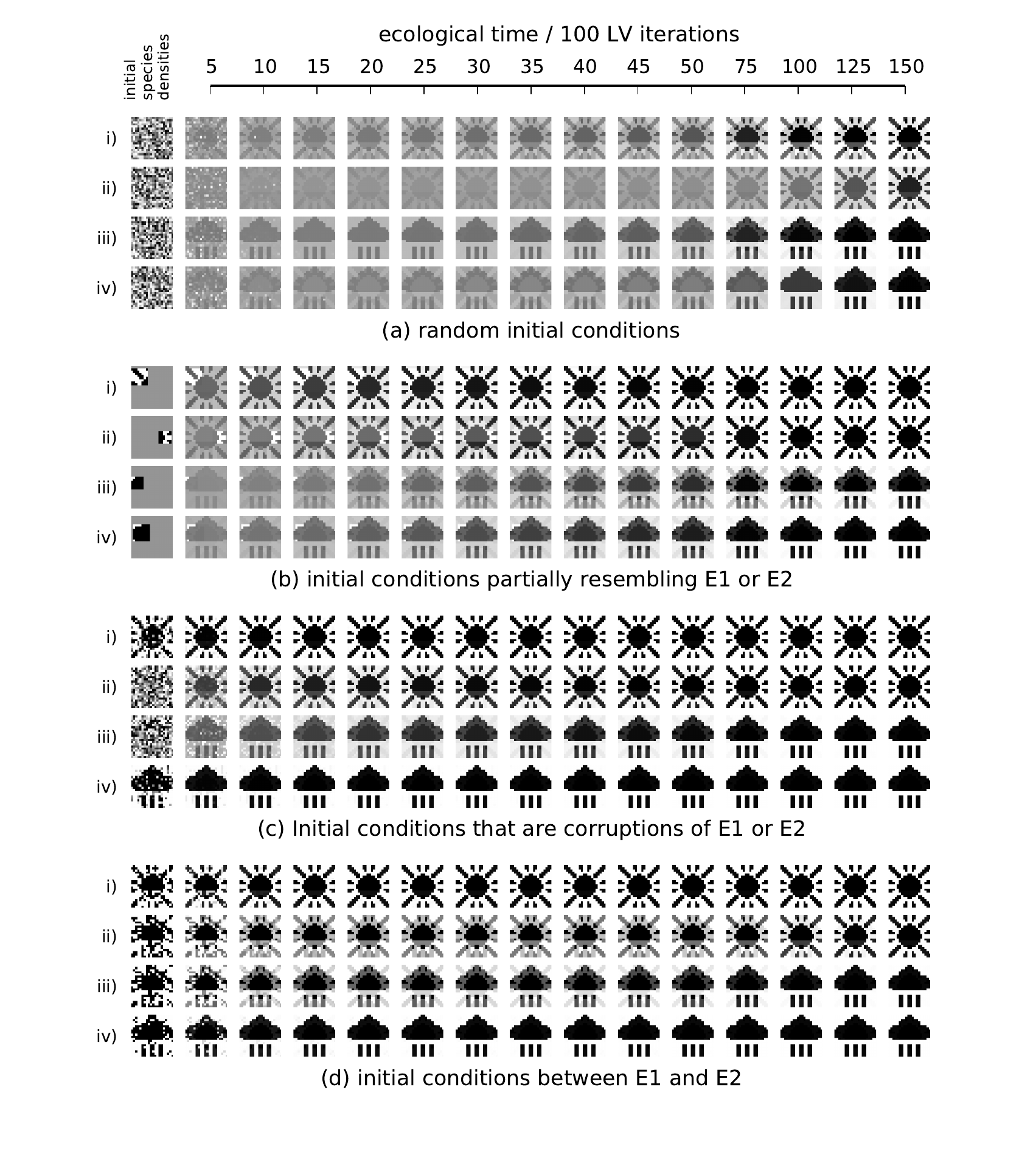}
\end{center}
  \caption{\textbf{Ecological dynamics after evolution in varying environmental conditions. }
      The evolved ecosystem exhibits two attractor states (rightmost frames) that are reached from various initial species densities (leftmost frames). 
      a) Random initial species densities develop into one of two possible attractors corresponding to the patterns of forcing experienced in the evolutionary past. 
      b) Initial configurations that resemble a small part of $E_1$ (i and ii) or $E_2$ (iii and iv) develop into equilibria that fully recreate $E_1$ and $E_2$ respectively. 
      c) Initial configurations that are partially randomised versions of $E_1$ (i. 20\%, ii. 80\%) or $E_2$ (iii. 80\%, iv. 20\%) develop into equilibria that `repair' the corresponding state. 
      e) For initial conditions between $E_1$ and $E_2$, ($E_1:E_2$ ratio = i.80:20, ii.55:45, iii.45:55 iv.20:80) the dynamics `recognise' the pattern that is resembled most closely.}
       \label{fig: Experiment 2}
      \end{figure}

After evolution in the varying environment (Experiment 2) either of the two patterns can be completely recalled or assembled from an initial subset of species. That is, when the initial species densities have just a few species present in a density that matches one of the previous patterns, the ecological dynamics act to recreate the full pattern to which that `partial stimulus' belonged (Fig. \ref{fig: Experiment 2}.b). This experiment also reveals more about how the stability and resilience of the community is affected by the presence of multiple memories. When the initial conditions are `corrupted' versions of a previous pattern, the complete pattern is restored, even when the corruption is severe (Fig. \ref{fig: Experiment 2}.c.) (thus maintaining/re-creating the current ecological pattern), and when the initial species densities partially resemble both patterns, the population dynamics `break symmetry', causing all species to adopt the pattern to which the initial conditions are closest (thus `choosing' between two ecological states - not blending them).\\

\begin{figure}[hp]
\begin{center}
\includegraphics[width = .85\textwidth]{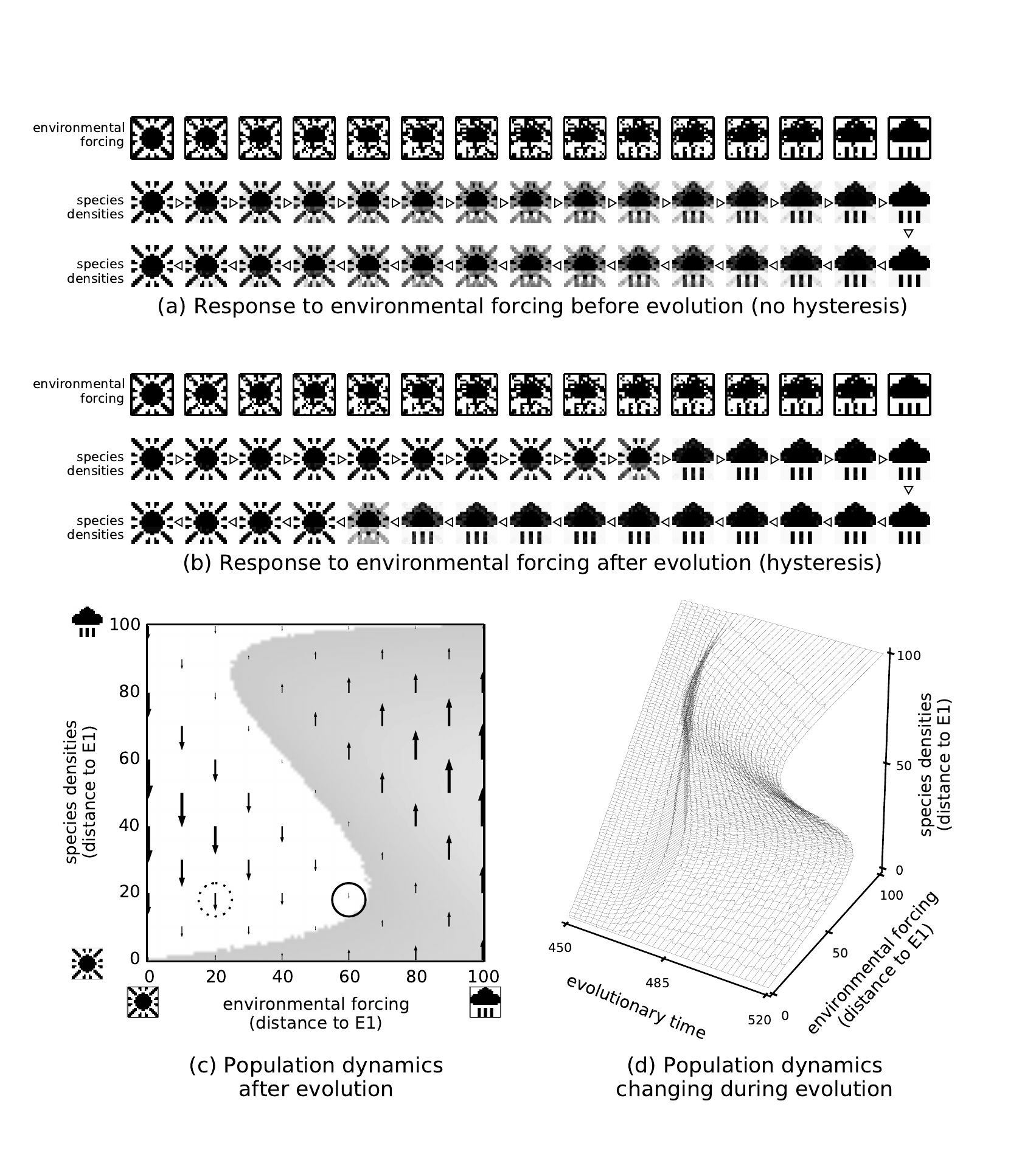}
\end{center}
  \caption{\textbf{Response to environmental forcing before and after evolution in varying environmental conditions.}
      a-b) Population dynamics under slowly-changing environmental forcing, changing first from $E_1$ to $E_2$ (middle row, left-to-right), and then back again from $E_2$ to $E_1$ (bottom row, right-to-left). 
      a) Before evolution of interactions, changes in species densities are proportional to forcing. 
      b) After evolution (Experiment 2), species densities show an abrupt switch between attractors with hysteresis. 
      c) Vector field for the population dynamics. The unstable equilibrium is revealed at the boundary of the shaded region indicating where species densities move away from $E_1$. Points near the critical transition (solid circle) have slower population dynamics than points far from critical transition (dashed circle). 
      d) Evolution of two-attractor system. Initially, change in species densities is proportional to environmental forcing. Around generation 470 non-linear but non-catastrophic transitions are observable. Finally, two stable attractors with a catastrophic transition and hysteresis.}
 \label{fig: Hysteresis}
\end{figure}

\subsection*{iv) Ecological memory can create multiple ecological stable states, and may exhibit critical transitions between them under subsequent environmental forcing.}

Fig. \ref{fig: Hysteresis}. examines the response of the ecosystem to patterns of environmental forcing that change linearly from $E_1$ to $E_2$. Before the evolution of interactions, the response of the ecosystem is proportional to the environmental forcing applied (Fig. \ref{fig: Hysteresis}.a). In contrast, after the evolution of interactions (Experiment 2), the response of the system is discrete or switch-like and exhibits significant hysteresis (Fig. \ref{fig: Hysteresis}.b). That is, as the pattern of environmental forcing moves incrementally from $E_1$ to $E_1$, the response of the system is to stay on $E_1$ considerably past the mid-point and then suddenly switch to $E_1$.  Conversely, when the environmental forcing is reversed, the ecosystem retains a configuration matching $E_1$ considerably past the mid-point before switching back to $E_1$. The dynamics underlying this hysteresis loop are shown by the vector field of species densities changes (Fig. \ref{fig: Hysteresis}.c, Fig. \ref{fig: Appendix response to forcing}.a). This also shows that the response of the population dynamics to perturbations in species densities slows down near the critical transition (consistent with early-warning signals for a tipping point  \cite{scheffer2009early,	dakos2008slowing}). Fig. \ref{fig: Hysteresis}.d. shows how the response of the ecosystem to forcing changes over evolutionary time. Around generation 470, the ecosystem exhibits non-linear but non-catastrophic transitions \cite{scheffer2001catastrophic}.  The catastrophic regime change is not a general instability property of the evolved system – it only occurs when the environmental forcing is similar to a past state that is remembered by the ecosystem – unstructured forcing results in a linear response (Appendix e). \\

These observations demonstrate a conversion of one type of ecological alternate stable state into another. Beisner \textit{et al.} \cite{beisner2003alternative}  describe the ``ecosystem" perspective on alternate stable states, which involves changes driven by abiotic environmental conditions, and the ``community" perspective, which involves multiple attractor states that can exist under fixed environmental conditions. Fig. \ref{fig: Hysteresis} shows a system that converts alternate ``ecosystem states" into alternate ``community states"; thus converting patterns from past environmental states into ecological memories. 

Fig. \ref{fig: Appendix bestiary} shows a `bestiary' of ecological attractors changing over evolutionary time, showing some diversity before settling down to the two-attractors shown in Figs. \ref{fig: Experiment 2} and \ref{fig: Hysteresis}. During long-term simulation we find that, as the forcing used to switch the system between attractors is of fixed value, while the effect of the evolved changes to interactions is ever-increasing, eventually the level of forcing applied is insufficient to shift the system from its current attractor. When this happens, the system becomes `stuck' at one attractor, effecting a breakdown in observed behaviour  (Appendix f, Figs. \ref{fig: Appendix bestiary} \& \ref{fig: Appendix Monte Carlo}).  

\section*{Discussion}
Our results formalise the intuitive idea that individual selection on ecological interactions produces positive feedback on species co-occurrence. By recognising an equivalence between this feedback and principles of unsupervised correlation learning we are able to use concepts from connectionist models to understand and illustrate the consequences of these changes for system-level behaviours. This makes several contributions to our understanding of evo-eco interactions:\\

Evo-eco dynamics have predictable consequences for collective behaviours.
The worked example developed in our simulations converts informal notions about the evolutionary and historical determinants of ecological processes into a model that makes specific predictions about how past ecological conditions alter the selective pressures on the component species and hence modifies their future ecological behaviours. This presents a specific model for non-trivial ecological memory that can be empirically tested (Appendix g). From this model we better-understand the necessary conditions for a distributed ecological memory to form, such as the presence of evolutionary trade-offs that cause species to become \textit{more dependent} on other species \cite{finn1976measures} rather than just becoming \textit{less competitive} with them (Appendix h). \\

Ecological communities can exhibit organised collective behaviours.
Under certain conditions, memories of past ecological states can be stored in a distributed way in the organisation of evolved ecological relationships. Such memories are not simply the summative or average result of multiple species each with individually alternate stable states. The connections that produce these behaviours are organised not by evolutionary adaptation at the community level, but rather by evolutionary adaptation at the individual level and `past experience' of historic environmental conditions. The organisation of the system is thus conditioned by past environmental conditions, causing it to collectively habituate to the patterns of perturbation it has experienced \cite{angeler2010identifying}.\\

Community assembly rules can be organised to re-assemble specific past states.
The assembly of complete and specific past ecological states may be triggered by partial environmental cues or a small number of founders (as in invasional meltdown) (Fig. \ref{fig: Experiment 2}.b) or, similarly, the system can recover each specific state from corruptions of that state (Fig. \ref{fig: Experiment 2}.c). The learned assembly rules result in a system that `classifies' initial compositions according to their similarity to past ecological states and will return community composition to the state that most closely resembles initial conditions (Fig. \ref{fig: Experiment 2}.d).  Ecological memories can thus direct subsequent community assembly to recreate multiple past ecological states in a complex and collective, but predictable, manner.\\

Stability and resilience tends to increase but instability and regime shifts are also predictable.
We find that evolutionary pressures on ecological interactions tend to increase ecosystem resilience (recovery after perturbations to species densities) (Fig. \ref{fig: experiment 1}) and adaptive capacity (robustness to environmental forcing) (Figs. \ref{fig: experiment 1} and \ref{fig: Hysteresis}) \cite{holling1973resilience,	gallopin2006linkages,	folke2006resilience,	cropp2002ecosystem, 	angeler2010identifying}. 	However, if the evolutionary history of an ecosystem has included a multi-modal distribution of environmental conditions, then this can result in alternative stable states (rather than universal stability) and may exhibit critical transitions in changing from one state to another \cite{holling1973resilience}. This switch-like change in the community (Fig. \ref{fig: Hysteresis}) is only exhibited when the forcing that is applied is similar to past forcing – when arbitrary forcing is applied the response may remain linear (Appendix e, Fig. \ref{fig: Appendix response to forcing}). Critical transitions between alternate stable states may thus reflect memories of specific past states and are not necessarily arbitrary non-linear responses to the current forcing pattern. Past experience of distinct environmental conditions (e.g. temperatures) may thus make future responses to related forcing (e.g. climate change) more likely to exhibit discrete changes in ecological states, critical transitions or tipping points \cite{higgins2002dynamics, 	scheffer2001catastrophic}.	This suggests that critical transitions are not necessarily the arbitrary failure of an ecological community but can be a matter of `recalling' alternate states familiar from past conditions.\\

Future work should explore the ultimate equilibrium of these evo-eco feedbacks (Appendix f, Figs. \ref{fig: Appendix bestiary} \& \ref{fig: Appendix Monte Carlo}), and investigate relaxation of some of the simplifying assumptions utilised in the memory behaviours illustrated here (Appendix h). In particular, this paper has not investigated the effect of evo-eco feedbacks on trophic (e.g. predator-prey) interactions or mutualisms, nor have we investigated the ecological analogues of other behaviours that can be produced by unsupervised learning in more general neural networks (e.g. with multi-layered or asymmetric connection structures). Some intriguing further research directions are also suggested:

Do brains learn in the same way that ecosystems evolve?
We have shown that ecosystems evolve in the same way that brains learn, but recognising evo-eco dynamics and connectionist learning models as different instantiations of the same underlying adaptive mechanisms also sheds light in the other direction, i.e. on cognitive processes   \cite{adams1998hebb,  	fernando2012selectionist}.

Can an ecosystem gain from experience? 
The idea of \textit{sequential selection}, where non-arbitrary organisation arises in a system over evolutionary time without selection being applied at the system level 
\cite{lenton2004clarifying,	betts2008second}	 suggests that a biological community ``may gain from experience" by using ``a system `memory' carried in the gene pool" \cite{lenton2002gaia}. 	Our work in other domains has shown that the more specific sense of system memory demonstrated in the current paper can improve the ability of an adaptive network \cite{gross2009adaptive}	 to solve constraint problems or optimise resource allocation problems (without an external reward signal) \cite{watson2010optimisation}.	  This suggests that adaptation at the ecosystem level is possible in a formal sense without group selection; not adaptation in a Darwinian sense, but rather in the same sense and by the same mechanism as connectionist models of organismic adaptation \cite{watson2011global}. 

Similarly, demonstration that ecosystems exhibit collective adaptive behaviours without being units of selection prompts inquiry as to whether these systems are capable of more complex computational tasks. For example, natural ecosystems are under very many constraints that limit species abundance (e.g. NCP availability). Does selection on individuals improve a system's ability to resolve these constraints? Hopfield networks are known to be able to solve complex constraint satisfaction problems \cite{hopfield1986computing}. Do these abilities translate to ecological networks? 

\section*{Conclusions}
We have introduced the framework of connectionist learning as a tool to expand our understanding of evo-eco dynamics and collective ecological behaviours. Within this framework we find that, despite not being an evolutionary unit, an ecological community can behave like an (unsupervised) learning system, creating internal organisations that collectively habituate to past environmental conditions, and actively recalling past responses to those conditions.

Previously there have only been two choices in how to interpret collective behaviours in ecosystems – i.e. either they have no system-level organisation or some mechanism of group selection must be involved. Our findings demonstrate that there is a third possibility. Ecological organisations that produce collective behaviours can arise from the positive feedback of individual natural selection and ecological population dynamics without invoking group selection.  Specifically, given the presence of evolutionary trade-offs, the effect of individual-level natural selection acting on interspecific relationships is dynamically equivalent to a mechanism of unsupervised correlation learning and ecosystems can thereby exhibit organised collective behaviours via the same principles of connectionist learning that apply to neural networks. What is it that ecosystems learn? We find that they have the potential to learn \textit{where to go} (i.e. evolved ecological attractors recreate past ecological states, where an attractor may be the climax community resulting from a successional process \cite{law1996permanence}), 	\textit{how to get there} (i.e. the successional or assembly process) and \textit{how to stay there} (i.e. the relationships that increase the resilience and stability of those mature ecological states). 	Of course, interpreting evo-eco dynamics as a connectionist learning system is not obligatory. A description in terms of individual natural selection and ecological population dynamics only is entirely compatible – indeed, we have provided this level of description for all the results in this paper. But recognising the equivalence with connectionist models enables us to convert and exploit well-understood concepts and results from this discipline to understand the organisation of ecological communities in new ways, and thereby to recognise the potential for predictable collective behaviours.


\section*{Appendices}
\subsection*{Appendix a: Individual selection in ecosystems}
A mutation to an individual in species $i$ that decreases the competitive effect, $\omega_{ij}$, of species $j$ on species $i$ directly affects the fitness of the individual carrying the mutation and not other individuals in species $i$, and can thus be favoured by individual selection. It is only changes to traits that directly affect the growth rate of an individual compared to the rest of the individuals in the species can be affected by individual-level selection. Traits that increase the growth rate of all individuals in the species equally have no differential individual benefit (despite conferring benefit to the species as a whole). In particular, a mutation to a trait in an individual in species $i$ that changes its competitive effect, $\omega_{ji}$, on some other species $j$, e.g. decreasing the density of a competitor species, may thereby indirectly increase the growth rate of species $i$. But this will benefit all individuals in species $i$, not just the mutant, and therefore has no differential selective benefit to the individual that bears the mutation \cite{wilson1980natural}.  Likewise, the competitive effect of species $j$ on species $i$ may, by virtue of normalising ecological constraints, be decreased as a side-effect of increasing the competitive effect of species $k$ on species $i$. But again this would not be favoured by individual selection as the benefit is felt by all individuals in species $i$  (conversely, changes to $\omega_{ij}$ could be selected under individual selection even though, as a result of indirect effects through changes in density of other species or through normalising ecological constraints, their net effect is to decrease the density of their own species). It is therefore only direct effects on individual fitness that are taken into account by the selection coefficient described here; i.e., Eq. 2. evaluates the change in growth rate of individuals in species $i$ due to changes in $\omega_{ij}$ and not $\omega_{ji}$, and furthermore, only changes to $\omega_{ij}$ caused by positive selection coefficients, not those caused by indirect normalisation effects. This correctly disregards any changes to a species growth rate that occurs as an indirect side-effect of altering the density of some other species. 

\subsection*{Appendix b: The relationship between rate of adaptation and product of species densities in more complex cases.}
In the main text, the rate of adaptation, $v_{ij}$, of each interspecific interaction coefficient is modelled with Eq.3 corresponding to the case where there is no interference between simultaneously segregating alleles at different loci. In large sexual populations with linked loci, the rate of adaptation will depend on the type of recombination, recombination rate, population size, the mutation rate and magnitude of mutations.  Here we compare the rate of adaptation of an interaction coefficient for three different models. In each case, the rate of adaptation, $v_{ij}$, of an interspecific interaction coefficient describing the fitness effect of species $j$ on species $i$, is $v_{ij} = x_i \mu \bar{P}$, where $x_i \mu$ is the rate with which beneficial mutations arise in species $i$, and $\bar{P}$ is the average probability that a single new mutation will ultimately fix (see main text). In all cases, $\bar{P}$ is a function of the selection coefficient $s_i = \frac{m_i}{ k_{ie}}  g x_j$  (Eq.2, main text) where $m_i$ is the intrinsic net growth rate of species $i$, $k_ie$ is the carrying capacity of species $i$ in environment $e$, and $g$ is the change in the interaction coefficient due to an individual mutation. Here we write $s_i = \beta x_j$, for clarity of the comparisons that follow.

\subsubsection*{Case a) No interference}

In simple cases when there is no interference between simultaneously segregating alleles at different loci (e.g. where genes are under weak selection per locus, free recombination and the linkage disequilibria among alleles sweeping to fixation are negligible), the probability of fixation, $\bar{P} =s_i$. Thus, as per Eq.3 main text:
\begin{equation}
v_{ij} = \beta \mu x_i x_j
\end{equation}
where $x_i$ is the density of species $i$, $x_j$ is the density of species $j$ and $\mu$ is the beneficial mutation rate.  

\subsubsection*{Case b) Linked genes on a linear genome}
Weissman \& Barton \cite{weissman2012limits} consider the effects of interference between linked genes on a linear genome. Here the genomic rate of fixation of beneficial mutations is (\cite{weissman2012limits} Equation 7):
\begin{equation}
v = \frac{v_0}{1+2 v_0 / R}
\end{equation}
where, $v_0$ is the genomic rate of fixation of beneficial mutations in the absence of interference and $R$ is the total genetic map length in Morgans. The authors use the approximation $v_0 = 2 x \mu s$, where $x$ is species density and $s$ is the selection coefficient. With $s_i = \beta x_j$ as before, this gives the rate of adaptation on an interaction:
\begin{equation}
v_ij = \frac{2 \beta \mu x_i x_j}{1 + 4 \beta \mu x_i x_j /R}
\end{equation}

\subsubsection*{Case c) Occasional outcrossing}
Neher\textit{ et al.} \cite{neher2010rate} study the rate of adaptation in unlinked loci in facultative sexuals where the rate of outcrossing is very small. Whereas Weissman and Barton examine the case of obligately sexual populations, this case represents occasionally/facultatively sexual populations (e.g. plants). On condition that $r^2 / s^2 \gg 4 x \mu$, the rate of accumulation of beneficial mutations in this case is given by (\cite{neher2010rate} Equation 12b):  
\begin{equation}
v \approx  x \mu s^2 \left( 
1 - \frac{4 x \mu s^2}{r^2}
\right)
\end{equation}
where $r$ is the outcrossing rate. With $s_i = \beta x_j$ as before, this gives the rate of adaptation on an interaction:
\begin{equation}
v_{ij} \approx x_i \mu (\beta x_j)^2 \left( 1 - \frac{4 x_i \mu (\beta x_j)^2}{r^2} \right)
\end{equation}

\subsubsection*{Comparison of the three cases}
       
      \begin{figure}[h!]
      \begin{center}
     \includegraphics[width = .9\textwidth]{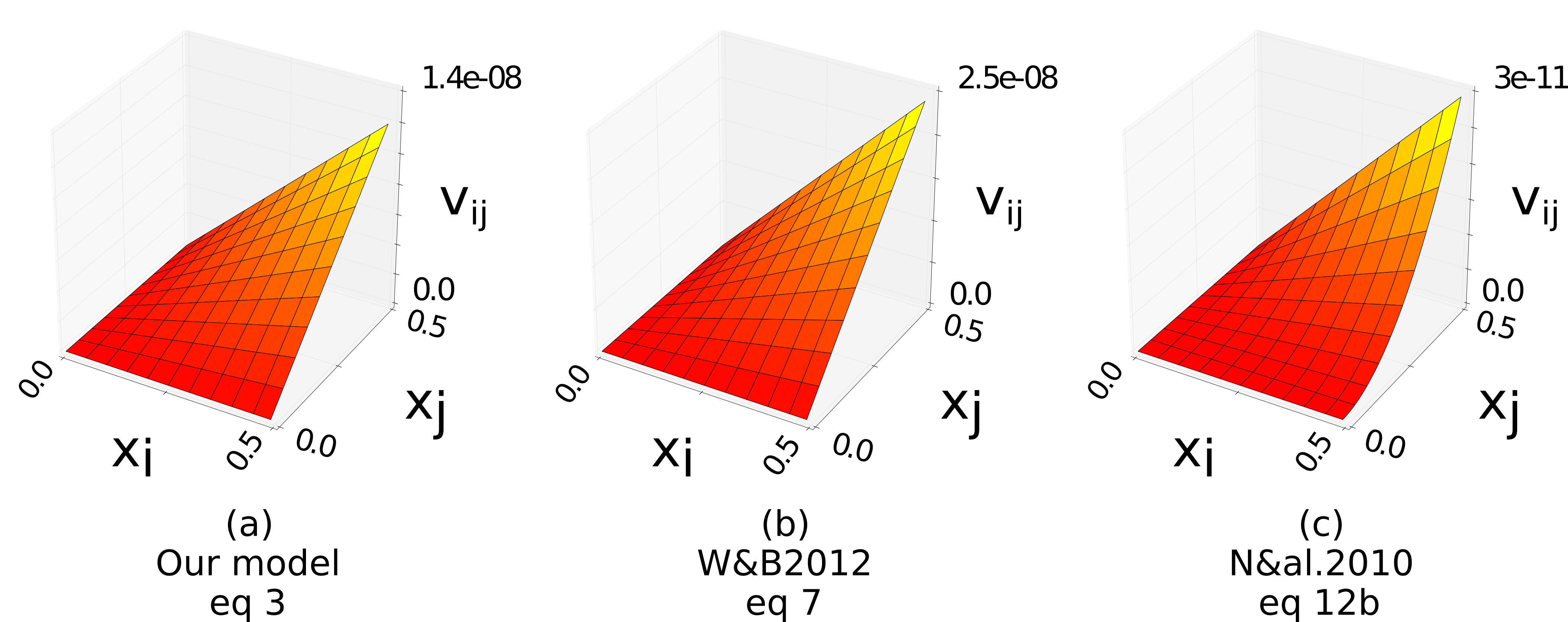} 
      \end{center}
  \caption{\textbf{Rate of adaptation $v_{ij}$ as a function of $x_i$ and $x_j$ for three different models.}
      a) Eq.3 from main text. b) Eq.7 from Weissman \& Barton 
\cite{weissman2012limits} , c) Eq.12b from Neher et al. 
\cite{neher2010rate}. We observe that (b) is very close to a linear scaling of (a) and, although (c) shows slight qualitative differences in the shape of the function, it maintains the essential qualitative characteristic. In all cases, the rate of evolutionary change in an interaction coefficient increases with the product of $x_i$ and $x_j$. (As per our simulation experiments: $k_{ie} = 10$, $m_i = 0.5$, $g = 0.1$, $\mu = 1.0 \times 10^{-5}$. For case b), the map length, $R = 1$. For case c), the out-crossing rate, $r = 0.01$.)}
\label{fig: Appendix: rates of adaptation}
 \end{figure}

Fig. \ref{fig: Appendix: rates of adaptation} plots the rate of adaptation $v_{ij}$ as a function of $x_i$ and $x_j$ for these three different cases. We observe that case a, where the rate of adaptation is directly proportional to the product $x_i x_j$ as modelled in our simulations, and the two more complex cases (b and c) are all qualitatively similar. Although in some cases the absolute rate of adaptation is more strongly limited by the recombination rate than the mutation supply or the strength of selection, for example \cite{weissman2012limits}, the relative rates of adaptation are still determined largely by the product of $x_i$ and $x_j$. More specifically, all three cases have the essential characteristic that the rate of adaptation is zero when either $x_i$ or $x_j$ is zero, and otherwise, the greater the value of one, the greater the rate of increase with the other. Thus, although the shape of the alternate functions differs from ours, the essential behaviour is preserved. Intuitively, mutations must be both created and selected for an interaction coefficient to evolve.\\
 

\subsection*{Appendix c: Additional methods: Normalisation, variable environments, measuring ecological attractors and model parameters}
\subsubsection*{Normalisation methods}
In each evolutionary step, all interaction terms in $\Omega (t)$ are updated by natural selection according to Eq. 3 to produce $\Omega ' (t)$  and then renormalised to produce $\Omega (t+1)$. 
Renormalisation preserves the conditions that for each species $i$ and all other species $j (j \ne i)$, 
$\sum_{j=1}^N \omega_{ij} (t) = Q_i$, and
$\sum_{j=1}^N \omega_{ji} (t) = Q_i$, where $Q_i < 0$ is a constant for each species. Specifically, an iterative row and column normalisation (below) is applied to $M(k=0) = \Omega'$, until the values of $M$ converge within a specified accuracy, i.e. $(\sum_{ij} ( m_{ij} (k+1) - m_{ij} (k) )^2 < 10^{-5}$, where $k$ is the iteration counter, as follows:
\begin{equation}
M(k+1)=\mbox{column\_norm} (\mbox{row\_norm} (M(k)))
\end{equation}
where row\_norm($m_{ii}$) = $m_{ii}$, column\_norm($m_{ii}$) = $m_{ii}$, i.e. self-interactions are unaffected, and
\begin{equation}
\mbox{row\_norm}(m_{ij(i \ne j}) = \frac{m_{ij}(k)}{\sum_{j=1 (j \ne i)}^N m_{ij} (k)}
\end{equation} and
\begin{equation}
\mbox{column\_norm}(m_{ij(i \ne j}) = \frac{m_{ij}(k)}{\sum_{i=1 (i \ne j)}^N m_{ij} (k)}
\end{equation}

\subsubsection*{Variable environments}
We investigate the effect of variable environments as follows. The carrying capacity of the $i^{th}$ species in a default ecological environment, $E_0$, is $k_{i0}$. For simplicity in our simulations we let $k_{i0}=k_0$, for all $i$, where $k_0$ is a constant. Prior to the evolution of interactions, this causes all species to equilibrate at the same density. To model the evolution of an ecosystem under varying environmental conditions that force or drive the ecosystem to adopt different ecological states, we define two other environmental conditions that alter carrying capacities. The pattern of equilibrium species densities under one environmental condition, $E_1$, increases the carrying capacity of some species to $k_0 + \alpha$ and decreases others to $k_0 - \alpha$, where $\alpha = 0.1$. In $E_2$, a different subset of species is increased/decreased in a similar manner. See Fig. \ref{fig: pictograms} main text.

\subsubsection*{Measuring ecological attractors}
We examine the ecological attractors in the ecosystem by Monte Carlo sampling, i.e., by repeatedly setting the species densities to random initial conditions and running to an equilibrium. To measure the inherent attractors induced by evolutionary changes, this sampling is carried out in the absence of environmental forcing – i.e., in $E_0$. In some experiments we also investigate the amount of environmental forcing required to push the ecosystem out of equilibrium in one pattern of species densities and into the attractor basin of another stable equilibrium.  Whenever, as here, interactions control the correlation of species densities and not their absolute densities, the complement of any attractor pattern is also necessarily an attractor \cite{hopfield1982neural, 	watson2010optimisation, 	watson2014evolution}. However, so long as initial conditions are more similar to the past states experienced during evolution than the opposite of those past states these unnatural attractors are precluded. Accordingly, we examine initial conditions, $x$, satisfying the condition $(|x-E_1|<|x-E_1' |)$ and $(|x-E_2|<|x-E_2' |)$ where $E'$ is the inverse of $E$ (i.e. $E'$ = $2 \bar{E}-E)$.

\subsubsection*{Model parameters}
$N=400$, number of species.\\
$m_i= 0.5$, growth rate of all species.\\
$s(t=0) = 0.1$, initial species densities.\\
$k_0=10$, a parameter governing the extrinsic component of carrying capacity in $E_0$. \\
$\alpha=0.1$ increment/decrement of particular carrying capacities in environments $E_1$ and $E_2$.\\ 
$T=1$, number of evolutionary changes applied in each environment before switching.\\
$\tau =5000$, number of ecological timesteps (Eq. 1) between `initial' and `final'.\\
$g=0.01$, constant of proportionality in selection-limited evolution (Eq. 3)\\
Interaction coefficients are initialised as follows:\\
\begin{equation*}
    \omega_{ij}(t=0) = \begin{cases}
               -1,                & \text{if  } i=j \text{ (i.e. self interactions}) \\
               -0.2,				& \text{otherwise}
           \end{cases}
\end{equation*}
$Q_i = \sum_{j=1 (j \ne i)}^N \omega_{ij}(t=0)$, normalisation constant (the sum of the non-self interactions in any one row/column remains equal to their sum at time $t=0$).

The quantitative values of these parameters will naturally have quantitative effects on the behaviour of the eco-evolutionary dynamics that we simulate.  Since the simulations are a phenomenological model of ecosystem evolution, what matters is the relative rather than absolute rates of adaptation on different interaction coefficients – in particular, which interactions increase, which decrease and which remain largely unchanged. This pattern, and its sensitivity to different modelling choices, is investigated in Appendix d.

\subsection*{Appendix d: Equivalence of Hebbian and evolved changes in more complex cases}
      \begin{figure}[h!]
      \begin{center}
  \includegraphics[width = \textwidth]{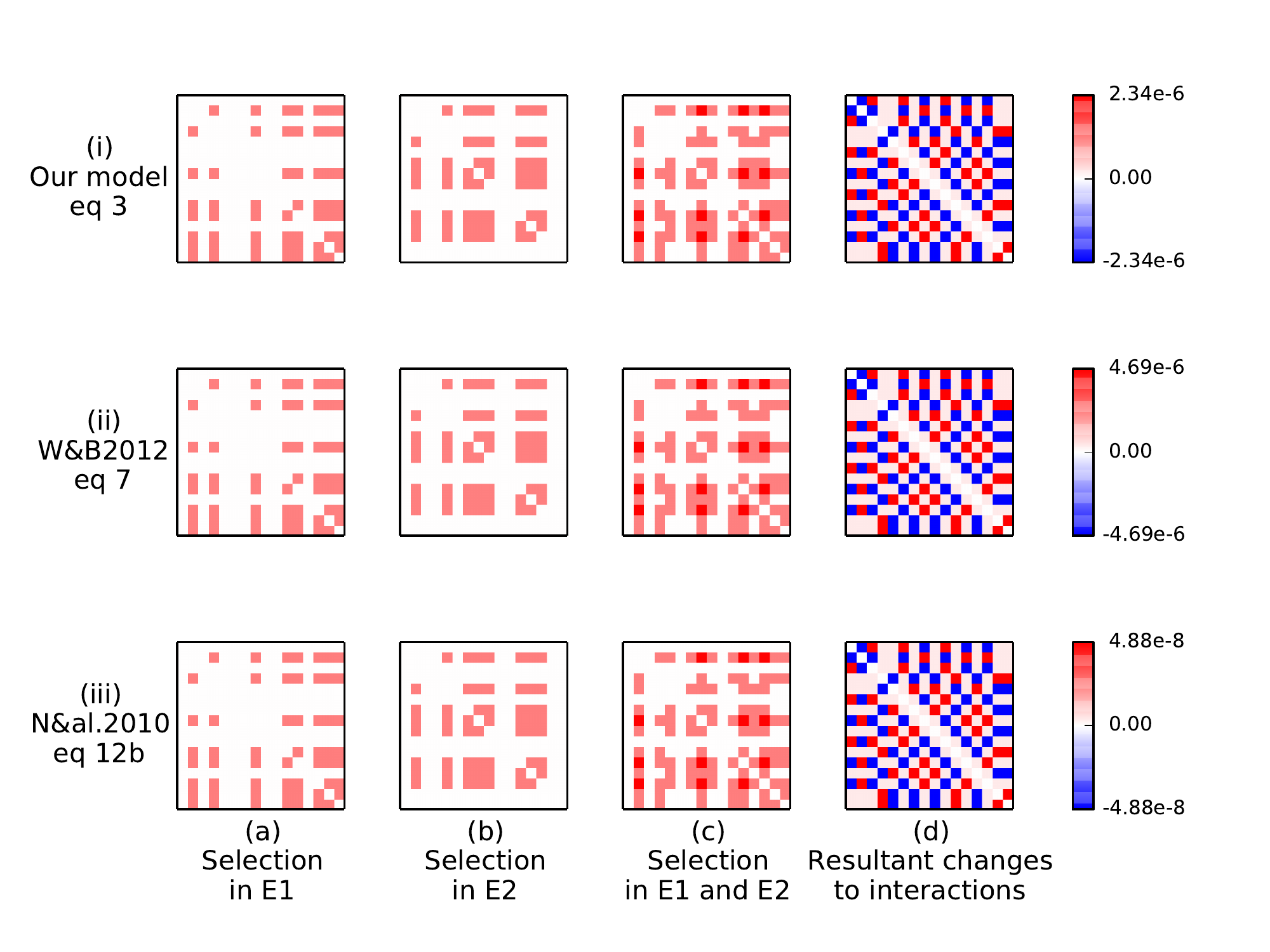}
\end{center}
  \caption{\textbf{Evolved changes to interactions are Hebbian in more complex population conditions.}
      Change in interactions between the first 16 species are shown under evolution in a changing environment. Rate of adaptation is controlled by our equation (top row), that from Weissman \& Barton 
      \cite{weissman2012limits}
      (middle row) and Neher et al. 
   \cite{neher2010rate}   
      (bottom row). a-b) the change in interactions due to direct selection effects (see Fig. \ref{fig: Hebbian equivalence} main text). d) When normalising ecological constraints are taken into account, some interactions are decreased, some left unchanged, and others are increased. The resulting direction of change is the same in all three cases and identical to Hebbian changes (Fig. \ref{fig: Hebbian equivalence}.e. main text). ($k_{ie} = 10$, $m_i = 0.5$, $g = 0.1$, $\mu = 1.0 \times 10^{-5}$, $\alpha = 3.5$. For case ii, the map length, $R = 1$. For case iii, the out-crossing rate, $r = 0.01$). For visualisation, the magnitude of changes in (d) are multiplied by 5.}
\label{fig: Appendix Hebbian equivalence}
      \end{figure}
In the main text the rate of adaptation of each interspecific interaction coefficient is modelled with Eq.3 corresponding to the case where there is no interference between simultaneously segregating alleles at different loci. Appendix b shows that the characteristics of the rate of adaptation in more complex cases is qualitatively similar although they are quantitatively different. Here we simulate evolution using these alternative models and incorporating normalising ecological constraints. Fig. \ref{fig: Appendix Hebbian equivalence} shows that the quantitative differences in the three equations do not alter the pattern of positive, negative and neutral changes that are produced in the evolving interaction matrix. Specifically, the pattern of changes in interactions have the same direction as the Hebbian model in all cases. Accordingly, there will be parameter ranges where they produce the same distributed memory phenomena in the ecosystem. Investigations of quantitative differences remain for future work.


\subsection*{Appendix e: Response to environmental forcing that is \textit{not} similar to environments experienced during evolution.}

\begin{figure}[h!]
\begin{center}
 \includegraphics[width = \textwidth]{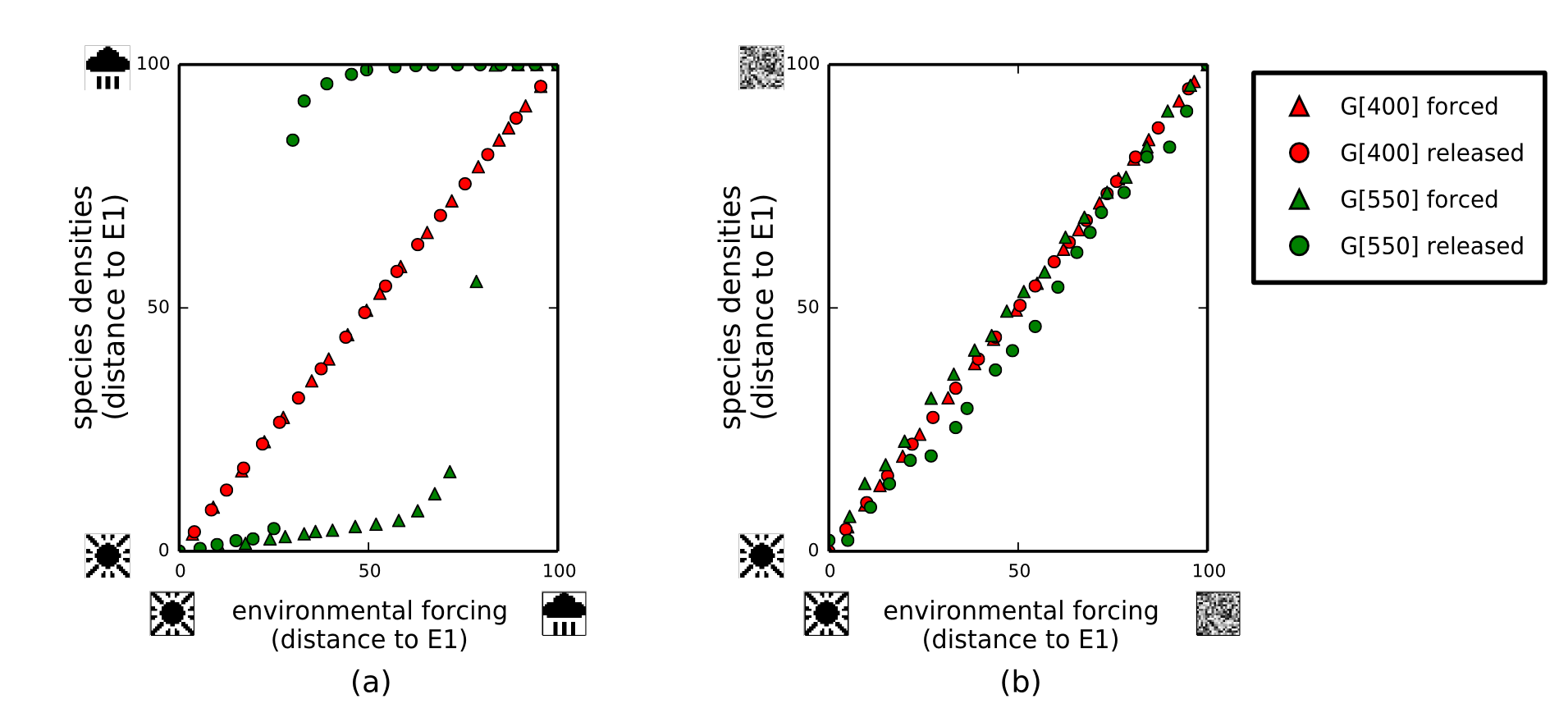}
\end{center}
  \caption{\textbf{Response to environmental forcing in different directions. }
      a) Environmental forcing that is similar to environments experienced during evolution (i.e. toward $E_2$, see thumbnail pictogram), b) Environmental forcing that is not similar to environments experienced during evolution (i.e. toward an arbitrary pattern, see thumbnail pictogram). }
\label{fig: Appendix response to forcing}
      \end{figure}

Fig. \ref{fig: Appendix response to forcing} shows that an ecosystem can exhibit a non-catastrophic response when forced in arbitrary directions (b) and simultaneously exhibit hysteresis and catastrophic regime shifts when forced in directions that have been experienced previously over evolutionary time (a). This emphasises that the evolved ecological memory causing the switching behaviour is conditioned by the systems' evolutionary history, and thus causes recall (or recognition) of a specific point in a multi-dimensional space of species densities, rather than a general stability/instability property resulting from unorganised or arbitrary evolutionary changes.\\


\begin{figure}[hp]
\begin{center}
\includegraphics[width = .9\textwidth]{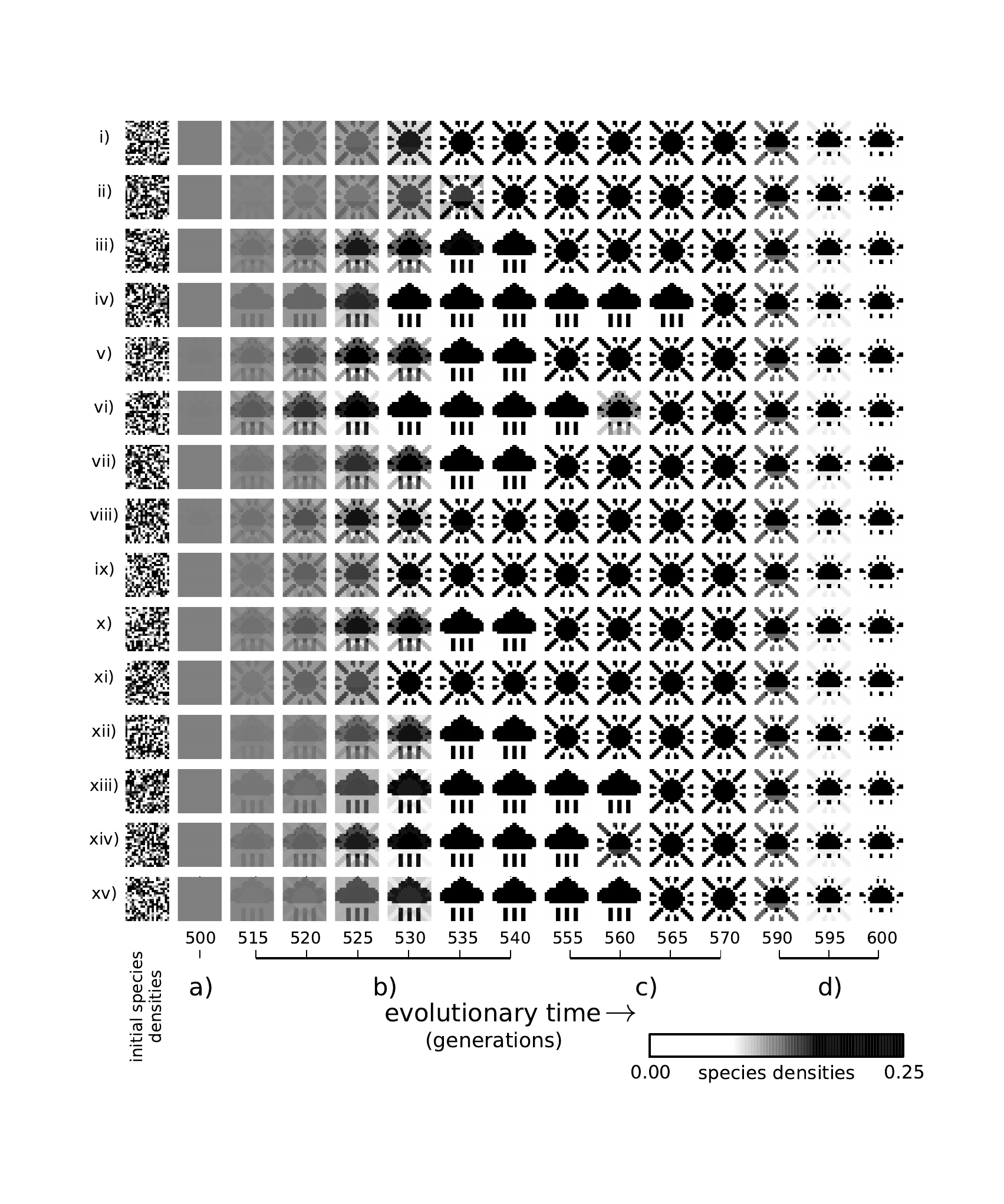}
\end{center}
  \caption{\textbf{`Bestiary' of ecological attractors changing over evolutionary time.}
      From an array of different random initial species densities (left-most column), the ecological states reached in the population dynamics changes over evolutionary time. 
      a) Initially, all initial conditions lead to the same ecological attractor (with all species at the same carrying capacity). 
      b) New attractor states begin to appear and become established. 
      c) In the long term, the two-attractor state is unstable and positive feedback causes one of the attractor states to `out-compete' the other. 
      d) Eventually the one remaining attractor breaks down as only the strongest species (those that were high density in both patterns) take over 
\cite{poderoso2007model}.
}
 \label{fig: Appendix bestiary}
      \end{figure}

\subsection*{Appendix f: Development and breakdown of multiple attractors over long evolutionary timescales.}

\begin{figure}[h!]
\begin{center}
 \includegraphics[width = \textwidth]{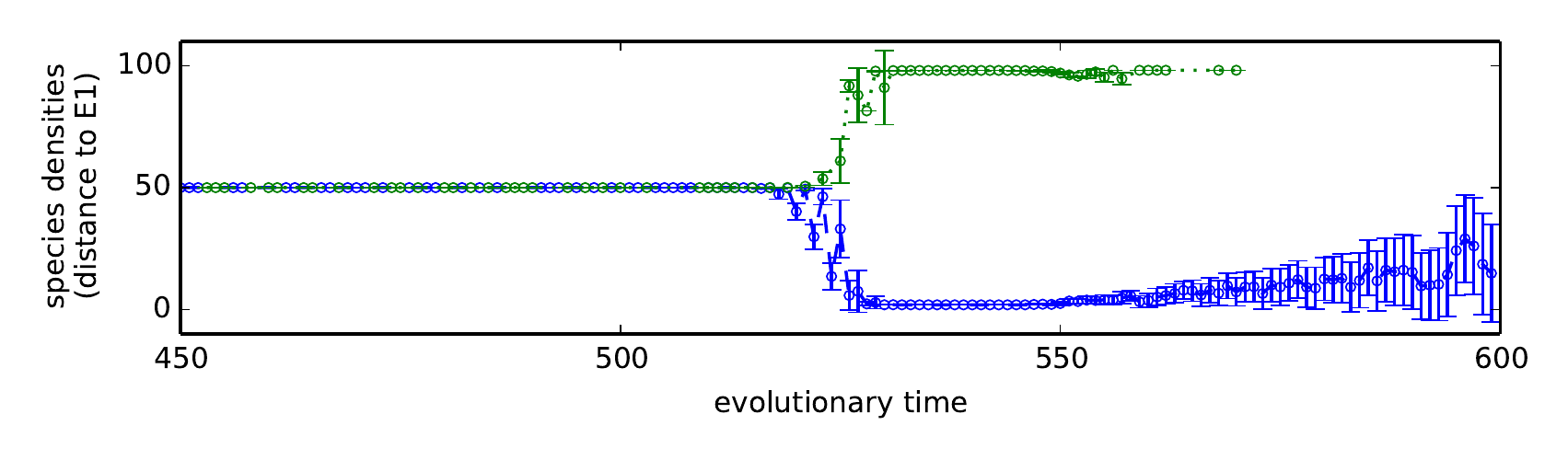}
\end{center}
  \caption{\textbf{In the long term the two-attractor system breaks down.}
      Monte Carlo sampling of the ecological attractor states from random initial species densities during evolutionary time. Initially, all attractor states contain species densities that are only minor deviations from the default attractor ($E_0$) in Euclidean distance. The signed pattern of the attractor state, i.e. in terms of $+/-$ with respect to the mean species densities, either matches $E_1$ (blue) or $E_2$ (green). As the two-attractor state emerges, at around generation 525 (a classic pitchfork bifurcation, but the unstable fixed point is not shown), the magnitudes (as well as signs) of the attractor states closely match the two targets. In the long term, one of the attractors, in this case $E_1$, outcompetes the other and becomes the only attractor. Eventually (after $\sim$575 generations), this attractor also degrades, i.e. the equilibrium magnitudes no longer match the original target closely (Fig. \ref{fig: Appendix bestiary}). }
 \label{fig: Appendix Monte Carlo}
      \end{figure}

Fig. \ref{fig: Appendix bestiary} shows how the attractors of the ecosystem change over evolutionary time in Experiment 2. 
Interestingly, we see that in the long term the two-attractor state is unstable because, rather than reinforcing the ecological patterns that are `forced' by the external environment, the system begins to reinforce its own patterns of behaviour \cite{watson2010optimisation},	and positive feedback causes one (slightly stronger) attractor to outcompete the other (Fig. \ref{fig: Appendix Monte Carlo}).\\


\subsection*{Appendix g: Empirical tests for distributed learning in ecosystems}

The dynamical behaviours we observe in the evolved ecosystem are consistent with ecological memory, alternate ecological states, succession dynamics, assembly rules, regime changes and founder effects observed in natural ecosystems. These behaviours follow from simple component principles (i.e. the availability of heritable variation in inter-specific interactions, and the presence of ecological constraints or evolutionary trade-offs) and direct evidence for these behaviours is testable. For example, consider the evolution of a small microbial community. Given a culturable community with stable coexistence dynamics, we could first test whether it has i) one or ii) alternative stable states. This requires sampling many different initial species compositions and allowing species densities to equilibrate. i) If a single state, we can then force the system into a different state (`alternate ecosystem state', \cite{beisner2003alternative})	 – e.g. by changing temperature, nutrient influx – and hold it there for evolutionary time. Then remove the forcing and retest for multiple attractors (`alternate community states'). If a memory has been conditioned by this forcing then a new attractor will be exhibited. ii) If the system initially has more than one attractor state, then we can estimate the basin size for each attractor by counting the number of different initial conditions that arrive at one or the other. By leaving the system in one attractor over evolutionary time this should increase the relative basin size in proportion to the time spent in that attractor. Next we need to assess the extent to which such a memory is collective or merely the sum of individual memories. This can be done by swapping-in evolved species for species in the original community one-by-one and assessing the relative contribution of individual and collective genetic changes on the dynamical behaviour of the system.

\subsection*{Appendix h: Asymmetric interactions, the importance of normalising ecological constraints, and other future work }

One important aspect of evo-eco dynamics that is highlighted by this model is the importance of normalising ecological constraints or evolutionary trade-offs for collective behaviours. These constraints prevent a species A from benefiting from the presence of species B without also becoming dependent on B. That is, it is not just the case that A grows faster in the presence of B, but that A's growth is slower when B is absent. Under these conditions, changes to interactions do not merely increase the growth of each species in a manner that is sensitive to its ecological context, but more specifically, they modify \textit{correlations} between species densities. We assume in the present model that an adaptation that, for example, decreases the niche overlap with one species increasing the niche overlap with others. But the extent to which species evolve dependencies rather than just (context-sensitive) individual advantages in natural ecosystems is an empirical matter – and from this work we recognise it as a matter that is centrally important to the possibility of collective behaviours that are more than the sum of the individual behaviours. 

This paper has investigated only competitive interactions and has not investigated mutualistic interactions or asymmetric interactions such as characteristic of trophic, e.g. predator-prey, relationships. The observation that selected changes to interactions are Hebbian does not depend on them being symmetric (or competitive). That is, Eq. 3 is not sensitive to any assumptions about the initial values of interaction coefficients, e.g., whether $\omega_{ij}$ and $\omega_{ji}$ are equal or even have the same sign, and therefore applies to predator-prey relationships as well as symmetric competitive interactions. Eq. 3 also shows that the selective pressures on \textit{changes} to interactions are symmetric i.e., $\Delta \omega_{ij} = \Delta \omega_{ji}$ (except for the influence of individually-varying carrying capacities), so there is no systematic reason for interactions to become asymmetric over evolutionary time. In the examples investigated in this paper the interaction coefficients are initialised symmetrically and, accordingly, they remain approximately symmetric. The evolutionary model could be applied to asymmetric interactions, but asymmetric interactions introduce the possibility of non-fixed point attractors, e.g. cycles, that complicate the behaviour of the eco-evolutionary dynamics and their measurement considerably. (We note that where $\omega_{ij}$ and $\omega_{ji}$ differ, the addition of multiple symmetric changes through natural selection will make them less asymmetric over evolutionary time, i.e., bring the ratio of these terms closer to 1, and could evolve them to take the same sign even when they started out with opposite signs. This implies that the effect of evolutionary change would be to increase the stability of the ecological dynamics and reduce or remove chaotic or cyclic attractors over time).

We have assumed that each interaction coefficient is independently modifiable whereas in natural populations traits may affect many interactions simultaneously. Here we chose to investigate scenarios where none of the interaction coefficients reach zero or go positive (which is possible in principle despite the normalisation employed). The equations used exhibit unstable behaviour in this case and a different approach to modelling would be required to handle mutualistic interactions. In natural populations one member of a population can gain selective advantage by changing its relationship to other members of its own species, but our simulations have fixed self-interactions at -1 and have investigated only the evolution of interactions with members of other species.

A key technical distinction between the recent work on associative memory in gene networks \cite{watson2014evolution} and the models utilised here is that the Lotka-Volterra equations represent unsigned (positive) state variables, as is natural for species densities, rather than signed (positive and negative) state variables representing under- or over- expressed gene activity (compared to some normal level). Although it is possible and common to model interesting dynamical behaviours using either signed or unsigned state variables in neural networks, the use of unsigned variables means that Hebb's rule, or natural selection, will only alter interactions in one direction, i.e., the product $x_i x_j$ is always positive (although crucially it may have different magnitudes). The assumption of normalising constraints that cause some interactions to become more competitive as a side effect of others becoming less competitive is thus important to the results that we have shown.

 In particular, the assumption of these normalising constraints means that changes to interactions, although motivated by increases in individual growth rates, have the effect of (also) altering the \textit{dependency} of one species on another. That is, an individual cannot evolve in a manner that avoids competition with one species, $x$, without also making their growth more dependent on the presence of some other species, $y$. Without these constraints, the effect of unconstrained changes to interactions is to make high density species fitter in all conditions, rather than making them dependent on the simultaneous high density state of specific species (and hence less fit in some conditions). It is therefore important for future work to investigate how different ways of modelling such constraints impact the behaviours illustrated here. For example, rather than a Lotka-Volterra model, a stoichiometric model of species interactions may alleviate the need for an explicit normalisation mechanism.
 
Assuming that ecological dynamics (i.e., changes in species density) are much more rapid than evolutionary changes (i.e., genetic changes affecting the coefficients of inter-species fitness dependencies) \cite{roughgarden1979theory}, 	most evolution occurs whilst ecological dynamics are at or near equilibrium, as modelled here. The behaviour of evo-eco dynamics when these processes have more similar timescales \cite{turcotte2011impact}	deserves attention. However, the fact that we model varying ecological conditions, causing the ecosystem to visit more than one ecological equilibrium, means that the interaction of ecological and evolutionary dynamics is non-trivial even though their timescales are kept separate in our simulations (following \cite{levin2011evolution}). 	Moreover, any model assuming a single ecological attractor will overlook the interesting behaviours modelled here, regardless of whether the timescales are separated or similar.

\bibliographystyle{Science}
\bibliography{eLife_bibliography}      

\end{document}